\documentclass[graybox]{svmult}

\usepackage{mathptmx}       
\usepackage{helvet}         
\usepackage{courier}        
\usepackage{type1cm}        
\usepackage{makeidx}         
\usepackage{graphicx}        
\usepackage{multicol}        
\usepackage{rotating}
\usepackage{multirow}
\usepackage{bm}
\usepackage[bottom]{footmisc}

\usepackage{upgreek}

\usepackage{bm}
\usepackage{amsmath}
\usepackage{listings}
\usepackage{amssymb}
\usepackage{braket}
\usepackage{mathtools}
\usepackage{hyperref}
\usepackage{algorithm}

\usepackage{eurosym}
\usepackage{amstext}

\usepackage[noend]{algpseudocode}

\graphicspath{{./Figures/}}

\makeindex             

\begin{document}

\title*{Multi-layered Network Structure: Relationship Between Financial and Macroeconomic Dynamics}
\titlerunning{Multi-layered network structure} 
\author{Kiran Sharma, Anindya S. Chakrabarti and Anirban Chakraborti}
\institute{Kiran Sharma \at School of Computational and Integrative Sciences, Jawaharlal Nehru University, New Delhi, India. \email{kiransharma1187@gmail.com},
\and Anindya S. Chakrabarti \at Economics Area, Indian Institute of Management, Ahmedabad, India. \email{anindyac@iima.ac.in}
\and Anirban Chakraborti \at School of Computational and Integrative Sciences, Jawaharlal Nehru University, New Delhi, India. \email{anirban@jnu.ac.in}}
%
%
\maketitle

\abstract{We demonstrate using multi-layered networks, the existence of an empirical linkage between the dynamics of the financial network constructed from the market indices and the macroeconomic networks constructed from macroeconomic variables such as trade, foreign direct investments, etc. for several countries across the globe. The temporal scales of the dynamics of the financial variables and the macroeconomic fundamentals are very different, which make the empirical linkage even more interesting and significant. Also, we find that there exist in the respective networks, core-periphery structures (determined through centrality measures) that are composed of the similar set of countries -- a result that may be related through the `gravity model' of the country-level macroeconomic networks. Thus, from a multi-lateral openness perspective, we elucidate that for individual countries, larger trade connectivity is positively associated with higher financial return correlations. Furthermore, we show that the Economic Complexity Index and the equity markets have a positive relationship among themselves, as is the case for Gross Domestic Product. The data science methodology using network theory, coupled with standard econometric techniques constitute a new approach to studying multi-level economic phenomena in a comprehensive manner.}

\section{Introduction}
Financial networks are major vehicles for transmitting shocks across different economic entities, which lead to complex dynamics. A well known phenomenon is that financial variables are considerably more volatile than macroeconomic variables with a much higher frequency, whereas, macroeconomic variables tend to show a much slower dynamics. A simple inspection of data suggests, there is wide variation even in intra-day stock returns, whereas macroeconomic variables move by a perceptible magnitude only over quarters or years if not longer time horizon. Thus these two types of variables differ both in frequency as well as the magnitude of oscillation. A directly related observation is that the magnitude of fluctuations of the financial variables	often seems decoupled from the fluctuations in the underlying macroeconomic variables. This is formally known as the excess volatility puzzle. 
	In an aggregate sense, growth rates of macroeconomic entities like firm-size variables shows bi-exponential distributions \cite{Stanley_1998}. But the corresponding financial indices typically have a power law structure which indicates much wider dispersion than exponential distributions. Thus, although the financial indices should reflect movements in underlying macroeconomic factors, it seems unlikely that the dynamics of individual financial time series can be readily explained by the dynamics of the corresponding economic variable.
	
In this chapter, we follow a complementary approach. In the finance literature, researchers have focused on factor models to relate economic variables to financial ones. We propose in the following that rather than looking at the time-series properties, a more useful approach could be to analyze the cross-sectional variation in the return structure and to find if there is any macro variable that explains the variation.
In particular, we posit that the aggregate financial network across countries are in sync with the dynamics of underlying macroeconomic fundamentals.	
The main idea stems from the work of Sharma et al. \cite{Sharma_2017_b}, which showed that at the sectoral level, there is a one-to-one mapping between the economic size of the sectors and centrality in the corresponding financial network. We extrapolate that idea to the country level. The novelty of the present approach lies in two factors. First, the earlier paper considered economic size of sectors measured by three indices (total market capitalization, revenue and employment) to be the underlying factors. Here, at the country level we extend the analysis by constructing country-to-country macroeconomic networks which underlie the financial network. In particular, we analyze the foreign direct investment network and trade network. Thus, it allows us to actually create a multi-layered network \cite{Lee_2015,Lee_2016} rather than just focusing on the size effect. Second, given that the Gravity equations are good models to understand country-level macroeconomic networks (see e.g. the works\cite{Fagiolo_JEIC_13a,Fagiolo_JEIC_13b} among others), we have an explanatory model of the relationship of this multi-layered network through the gravity model. Wang et al. \cite{Wang_2018} constructed and analyzed a cross-country financial network. They analyzed the topological properties of the network with different clustering algorithms. We differ significantly from their approach with our emphasis on country-level fundamentals and their connections with the financial network. Main problem-wise the closest work to ours is of Qadan and Yagil \cite{Qadan_15}, who analyzed a very similar problem with econometric techniques. But they did not explicitly consider network topology. Hence, our results complement their findings. Finally, Bookstaber and Kenett \cite{Kenett_16} constructed a multi-layered map of the financial system and analyzed its topology. Our usage of multi-layered network was motivated by that paper, but our emphasis on the macroeconomic variables provide new and different features of the data.
	
The main points of this chapter are as follows: First, there is a relationship between centrality measures of financial return correlation network across countries and the same for trade and FDI networks. Second, from a multi-lateral openness perspective, we show that even for individual countries, larger trade connectivity is positively associated with higher financial return correlations.	Third, we analyze the network architecture by using different clustering algorithms, which in turn allows us to identify the countries that are at the core or at the periphery.

\section{Macroscopic View}
\label{subsec:macro}
In this section, we study the relationship between financial indices return network, trade and foreign direct investment (FDI) networks as a multiplex network for 18 European countries. Next, we study the world stock market and relationship of macro variables and indicators like economy size, Economic Complexity Index, etc. 
\subsection{Data Description} 
\label{subsec:data-desp}	
For the macro-level analysis, we have used the data of the adjusted closing price for 18 European countries downloaded from Thomson Reuters Eikon database \cite{Thompson_reuters}, within the time period of 2001-2009. The countries are:
(1) \textbf{AUT}- Austria 
(2) \textbf{BEL}- Belgium 
(3) \textbf{CZE}- Czech Republic 
(4) \textbf{DEU}- Germany 
(5) \textbf{DNK}- Denmark 
(6) \textbf{ESP}- Spain 
(7) \textbf{FRA}- France 
(8) \textbf{GBR}- United Kingdom
(9) \textbf{HUN}- Hungary 
(10) \textbf{IRL}- Ireland
(11) \textbf{ITA}- Italy  
(12) \textbf{LVA}- Latvia 
(13) \textbf{NLD}- The Netherlands 
(14) \textbf{POL}- Poland 
(15) \textbf{PRT}- Portugal 
(16) \textbf{ROU}- Romania 
(17) \textbf{SVK}- Slovak Republic
and
(18) \textbf{SWE}- Sweden.
{Data for Foreign Direct Investment (FDI) and international trade for same 18 European countries is downloaded from {\it External and intra-EU trade, A statistical yearbook}, 2011 edition published by {\it eurostat}. 
To study the evolution of world stock markets, we have used the adjusted closing price of 51 market indices across the globe downloaded from the Thomson Reuters Eikon database, within the time period of 2001-2015. 
The countries: 
(1) \textbf{USA}- The United States of America
(2) \textbf{CAN}- Canada
(3) \textbf{BRA}- Brazil
(4) \textbf{ARG}- Argentina
(5) \textbf{MEX}- Mexico
(6) \textbf{CHL}- Chile
(7) \textbf{VEN}- Venezuela
(8) \textbf{PER}- Peru
(9) \textbf{JPN}- Japan 
(10) \textbf{SGP}- Singapore
(11) \textbf{CHN}- China
(12) \textbf{AUS}- Australia
(13) \textbf{HKG}- Hong Kong
(14) \textbf{KOR}- Korea
(15) \textbf{IND}- India
(16) \textbf{IDN}- Indonesia
(17) \textbf{MYS}- Malaysia
(18) \textbf{THA}- Thailand
(19) \textbf{PHL}- Philippines
(20) \textbf{PAK}- Pakistan
(21) \textbf{LKA}- Sri Lanka
(22) \textbf{GBR}- United Kingdom
(23) \textbf{FRA}- France
(24) \textbf{ITA}- Italy
(25) \textbf{ESP}- Spain
(26) \textbf{RUS}- Russia
(27) \textbf{NLD}- The Netherlands
(28) \textbf{CHE}- Switzerland
(29) \textbf{SWE}- Sweden
(30) \textbf{POL}- Poland 
(31) \textbf{BEL}- Belgium
(32) \textbf{NOR}- Norway
(33) \textbf{AUT}- Austria
(34) \textbf{DNK}- Denmark
(35) \textbf{GRC}- Greece
(36) \textbf{PRT}- Portugal
(37) \textbf{HUN}- Hungary
(38) \textbf{IRL}- Ireland
(39) \textbf{TUR}- Turkey
(40) \textbf{ROU}- Romania
(41) \textbf{SVK}- Slovak Republic
(42) \textbf{HRV}- Croatia
(43) \textbf{CZE}- Czech Republic
(44) \textbf{LVA}- Latvia
(45) \textbf{DEU}- Germany
(46) \textbf{QAT}- Qatar
(47) \textbf{SAU}- Saudi Arabia
(48) \textbf{OMN}- Oman
(49) \textbf{KWT}- Kuwait
(50) \textbf{TUN}- Tunisia
and
(51) \textbf{ZAF}- South Africa, spread across the continent of Latin America, Asia, Europe and Africa.
Economic complexity data is downloaded from Atlas of Economic Complexity \cite{ECI} for the period 2001-2015.
GDP (per capita by country) data is downloaded from Knoema world data Atlas \cite{GDP}.
\subsection{Some Basic Measures}

We consider aggregate stock market indices $\{P_{it}\}_{i\in N,t\in T}$ for $N$ countries and $T$ periods. We construct the return series by taking simple log differences of the prices levels
\begin{equation}
\{r_{it}\}_{i\in N,t\in T-1} = \log (\{P_{it}\}_{i\in N,t\in T}/\{P_{i(t-1)}\}_{i\in N,t\in T}). 
\end{equation}
Next, we construct the correlation matrix $\rho_{N\times N}$ from the $N$ time-series.
We have used eigenvector centrality (we refer to the measure as EVC) 
to measure centrality of different nodes in a given network.
EVC is defined by a vector $e_{N\times 1}$ which solves 
\begin{equation}
\lambda e=\rho e,
\end{equation}
where $\lambda$ is an eigenvalue of the matrix $\rho$. EVC is defined as the eigenvector $e$ corresponding to 
\begin{equation}
\lambda=\max_{\{i \in N\}}  \{ \lambda_i \}.
\end{equation}
For all variables $x$, we construct the z-score of the same as 
\begin{equation}
x_z=(x-E(x))/\sigma_x.
\end{equation}

	\subsection{The Relationship Between Financial Indices, International Trade, and Foreign Direct Investment}
	\label{subsec:multiplex}
	In this section, we try to find the FDI-trade linkage between host and home countries  \cite{Liu_1998, Lee_2002}, and their effect on financial indices in the form of a multiplex network. Here, we show empirical evidence for 18 European countries (see data description in section \ref{subsec:data-desp}) whether financial indices and international trade of these nations are substitute or compliments, i.e., whether a great market index held by a nation is associated with decreases or increases of its export and imports. The effect of FDI on trade is always a concern for the policymakers. So we studied the effect of FDI on international trade and financial indices. For this analysis, we have chosen both the developed and developing countries of European continent. The literature on FDI and trade generally points to a positive growth relationship.
	
	\begin{figure}[]
		\centering
		\includegraphics[width=0.75\linewidth]{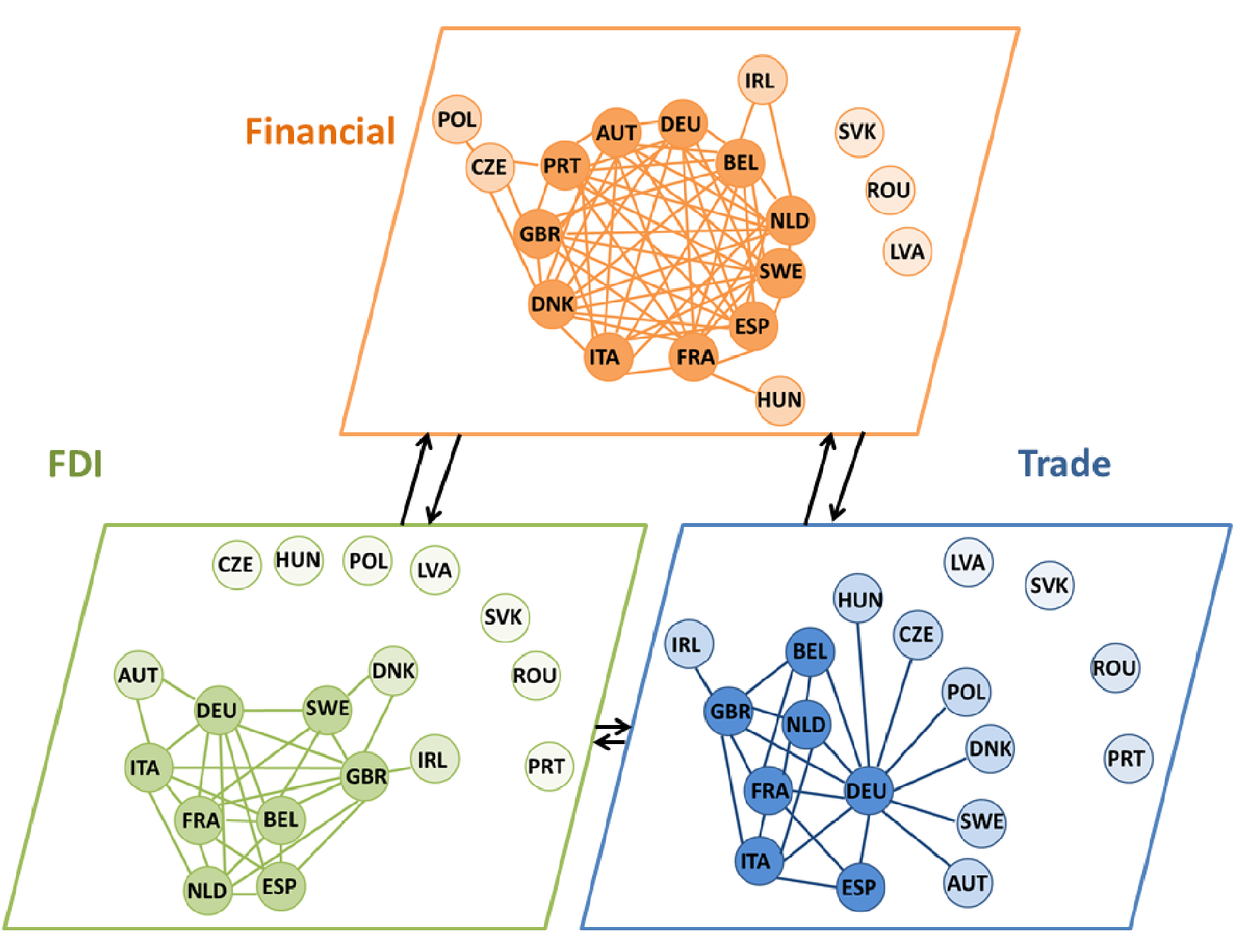}\llap{\parbox[b]{3.8in}{(a)\\\rule{0ex}{2.6in}}}\\
		\vspace{2em}
		\includegraphics[width=0.32\linewidth]{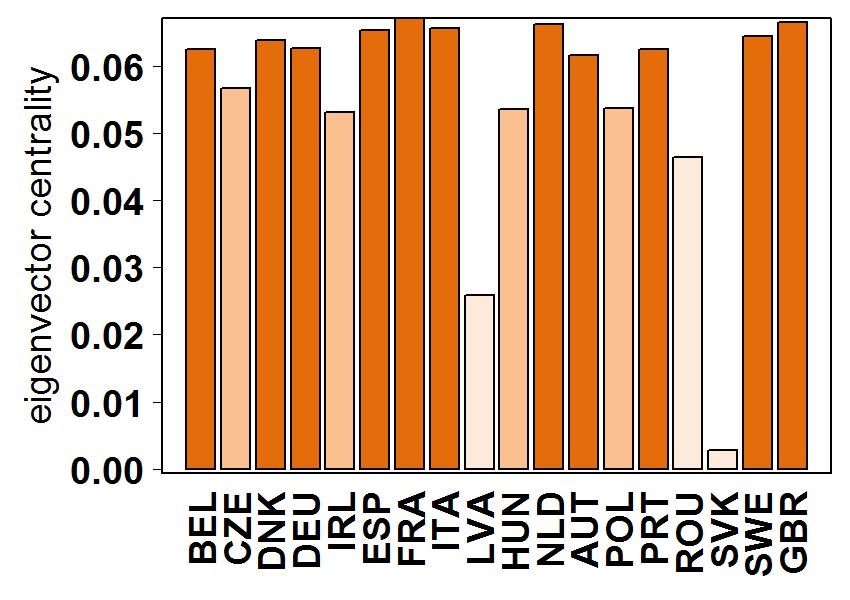}\llap{\parbox[b]{1.6in}{(b)\\\rule{0ex}{1.1in}}}
		\includegraphics[width=0.32\linewidth]{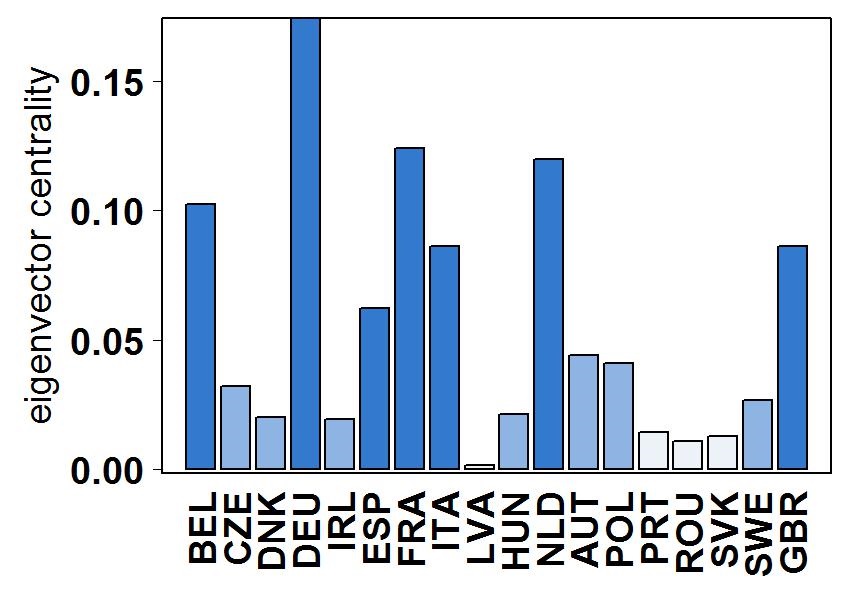}\llap{\parbox[b]{1.4in}{(c)\\\rule{0ex}{1.1in}}}
		\includegraphics[width=0.32\linewidth]{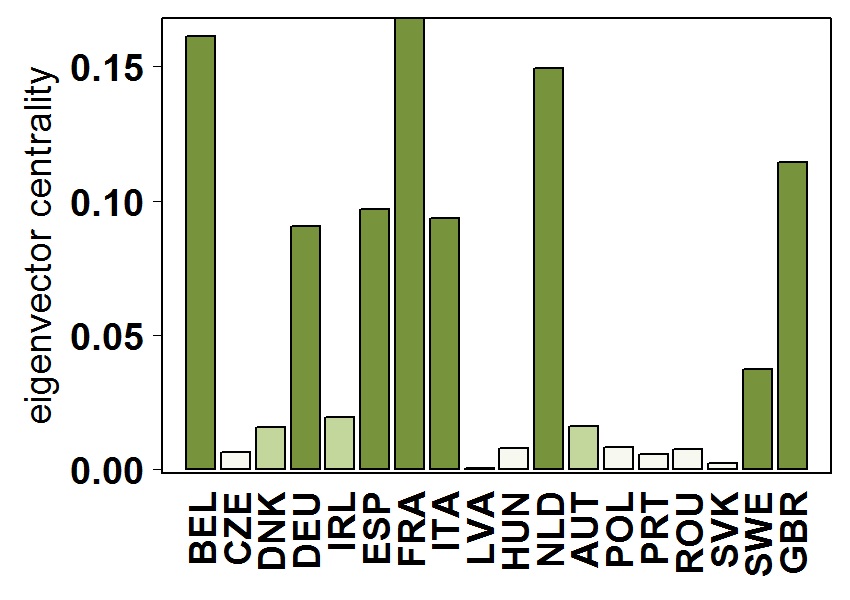}\llap{\parbox[b]{1.4in}{(d)\\\rule{0ex}{1.1in}}}
		\caption{(a) Multiplex network for 18 European countries for the year 2008. Financial indices are the top-level network and macroeconomic entities (Trade and FDI) are base-level networks. Eigenvector centrality for (b) Financial indices, (c) Trade and (d) Foreign direct investment (FDI). We divided the EVC in all the networks with three different shades (light to dark).}
		\label{fig:multiplex}
	\end{figure}
	In Fig. \ref{fig:multiplex}, we present a multi-layered network view of the 18 European countries.  
	In panel (a), we construct that the base-level networks formed across countries in terms of trade flow and FDI flow. Both of these two networks capture the connections through economic variables. The top-layer, on the other hand, has been constructed from the financial indices. Here, we examine the relationship between the upper layer of financial network and lower levels of FDI and trade networks. The countries occupying central positions in the correlation network are also central in the corresponding trade and FDI network.
	In panels (b), (c) and (d), we show the eigenvector centralities of the corresponding countries for these three variables. We cut down the EVC at three levels and that is reflected in the network of financial indices, trade and FDI. In all the networks three countries: SVK, ROU and LVA are forming no link with other countries. PRT is not forming any link in trade and FDI network. CZE, HUN, POL are not forming links in trade network with rest of the countries. Germany is Europe's one of the developed country and strongest economy due to its highly skilled labor force, high quality of life for its resident, etc. as visible in Fig. \ref{fig:multiplex}(a) trade network. 
	We computed the eigenvector centrality (normalized) of financial indices, international trade and FDI. Then, we regress these three variables as shown in Fig. \ref{fig:LR_FI_Trade_FDI}. EVC's of trade and FDI points to a positive growth relationship having $\beta=1.04\pm0.17$ with $p-value=0.00001$. Germany is an outlier. EVC's of trade and financial indices are also showing positive slope having $\beta=0.17\pm 0.07$ with $p-value=0.03$. Latvia and Slovakia are outliers. EVC's of FDI and financial indices are showing a mildly positive slope having $\beta=0.15\pm0.06$ with $p-value=0.03$.
	
	\begin{figure}[]
		\centering
		\includegraphics[width=0.32\linewidth]{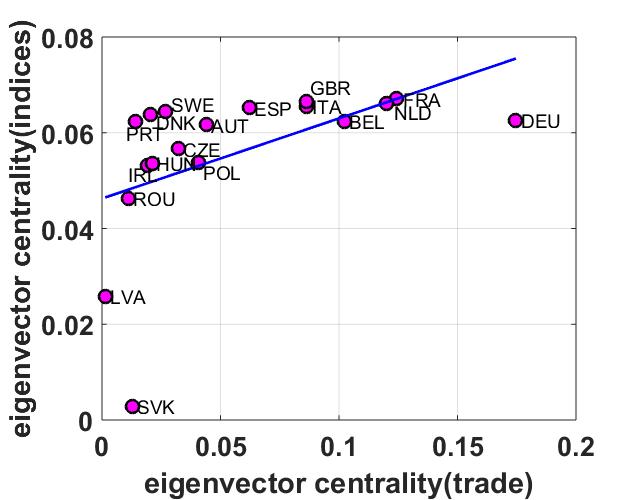}\llap{\parbox[b]{1.6in}{(a)\\\rule{0ex}{1.1in}}}
		\includegraphics[width=0.32\linewidth]{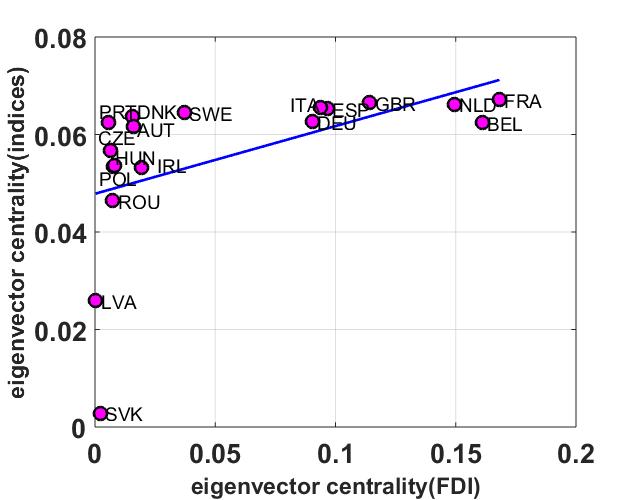}\llap{\parbox[b]{1.6in}{(b)\\\rule{0ex}{1.1in}}}
		\includegraphics[width=0.32\linewidth]{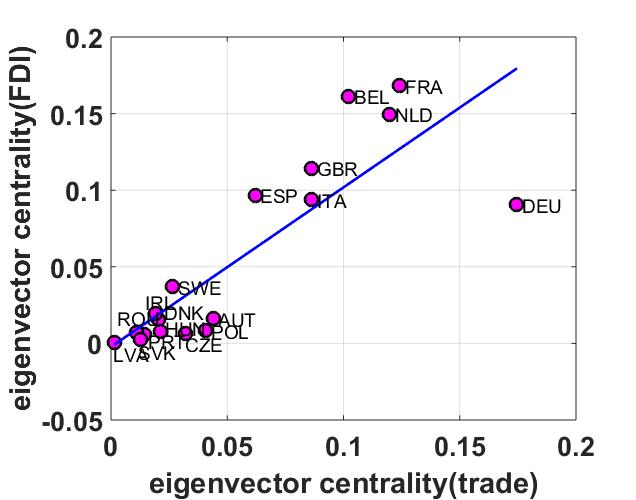}\llap{\parbox[b]{1.6in}{(c)\\\rule{0ex}{1.1in}}}
		\caption{Linear regression of normalized eigenvector centrality between (a) trade and financial indices having $\beta=0.17\pm 0.07$ with $p-value=0.03$,  (b) FDI and financial indices having $\beta=0.15\pm0.06$ with $p-value=0.03$, (c) Trade and FDI having $\beta=1.04\pm0.17$ with $p-value=0.00001$, of 18 European countries for year 2008.}
		\label{fig:LR_FI_Trade_FDI}
	\end{figure}
	To see the co-evolution of trade and financial indices, we regress the EVC's of indices and trade for the period 2003, 2005, 2007, and 2009 as shown in Fig. \ref{fig:Trade_Vs_Financial_Network}. The positive slopes of the best fit line indicates that higher centrality in the financial network is occupying more central positions in the trade network. This pattern holds true for all four time periods, both before and immediately after the financial crisis. Thus, we show that there exists a mapping between the financial network and the trade network. The co-movement of three countries: SVK, LVA and ROU is traced. In the year 2009, ROU came closer to the rest of the countries (as seen in Fig.  \ref{fig:Trade_Vs_Financial_Network}(h)). Germany is always an outlier.
	
	
	\begin{figure}[]
		\centering
		\includegraphics[width=0.42\linewidth]{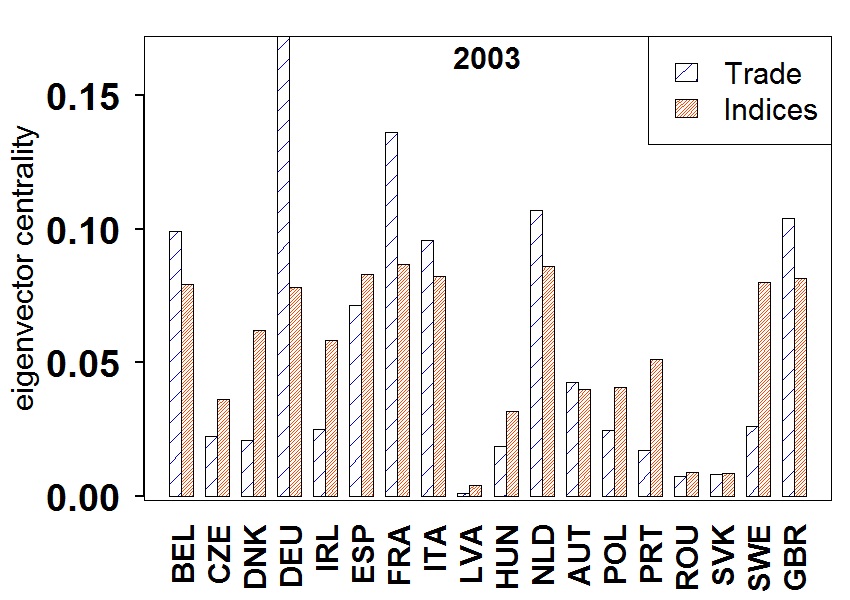}\llap{\parbox[b]{2.2in}{(a)\\\rule{0ex}{1.3in}}}
		\hspace{0.5em}
		\includegraphics[width=0.4\linewidth]{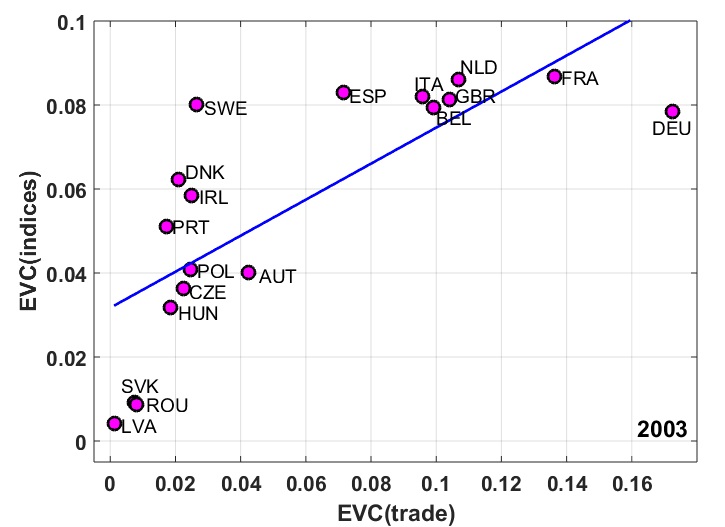}\llap{\parbox[b]{2in}{(e)\\\rule{0ex}{1.3in}}}\\
		\includegraphics[width=0.42\linewidth]{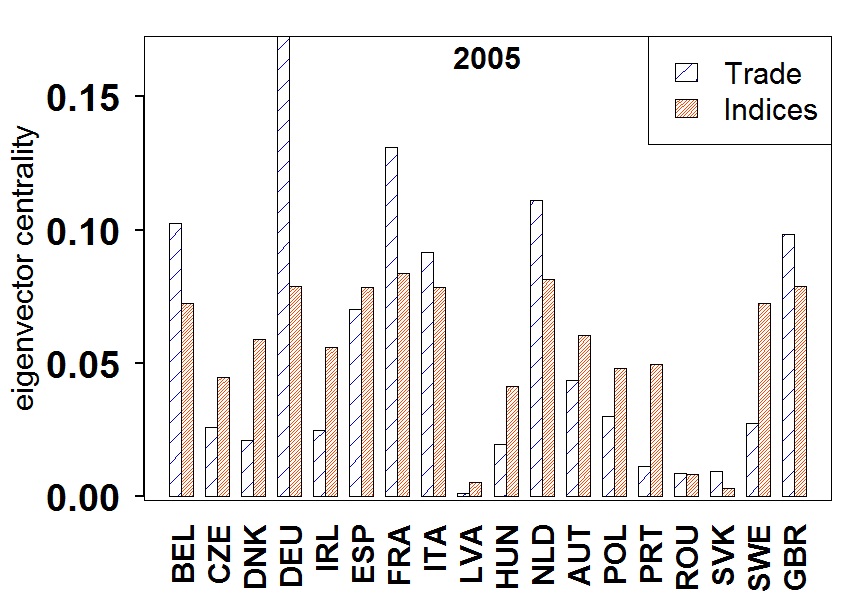}\llap{\parbox[b]{2.2in}{(b)\\\rule{0ex}{1.3in}}}
		\hspace{0.5em}
		\includegraphics[width=0.4\linewidth]{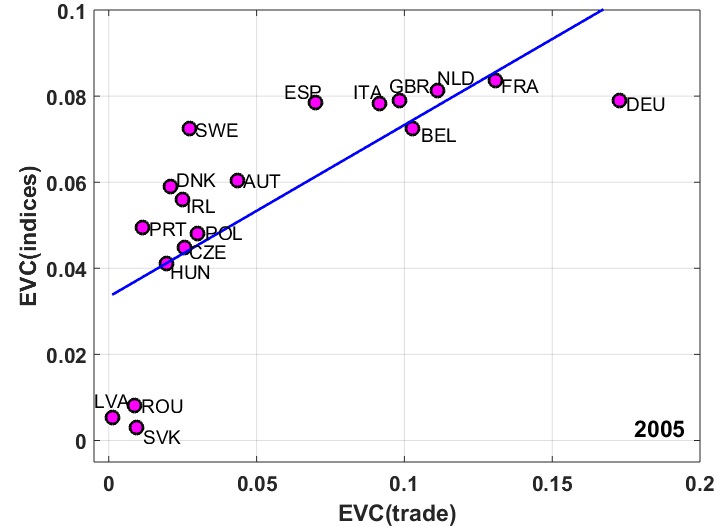}\llap{\parbox[b]{2in}{(f)\\\rule{0ex}{1.3in}}}\\
		\includegraphics[width=0.42\linewidth]{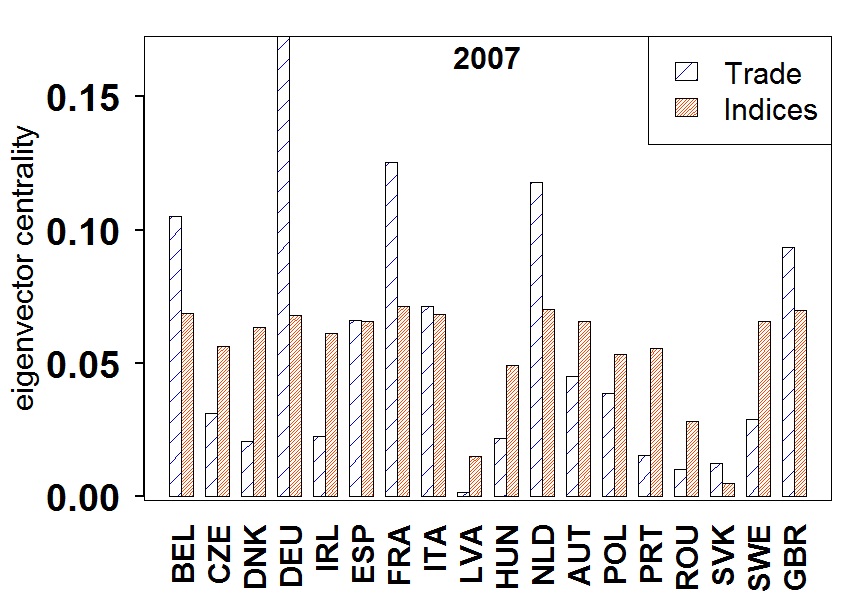}\llap{\parbox[b]{2.2in}{(c)\\\rule{0ex}{1.3in}}}
		\hspace{0.5em}
		\includegraphics[width=0.4\linewidth]{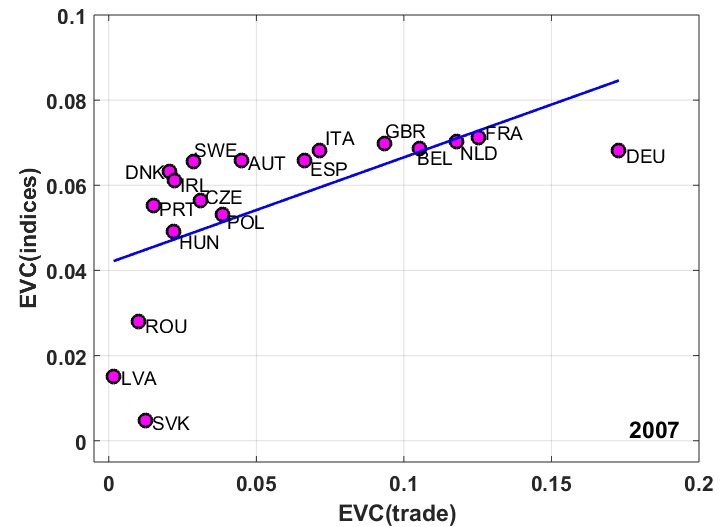}\llap{\parbox[b]{2in}{(g)\\\rule{0ex}{1.3in}}}\\
		\includegraphics[width=0.42\linewidth]{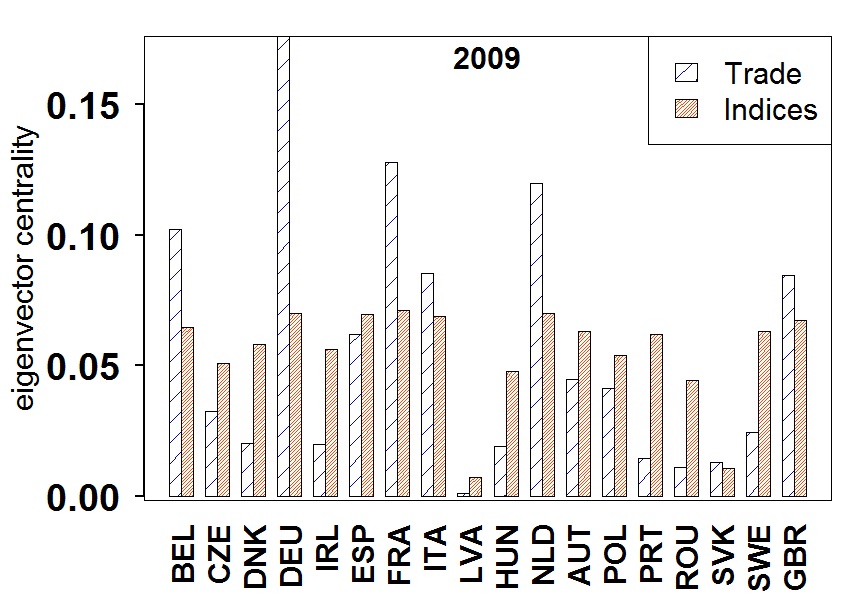}\llap{\parbox[b]{2.2in}{(d)\\\rule{0ex}{1.3in}}}
		\hspace{0.5em}
		\includegraphics[width=0.4\linewidth]{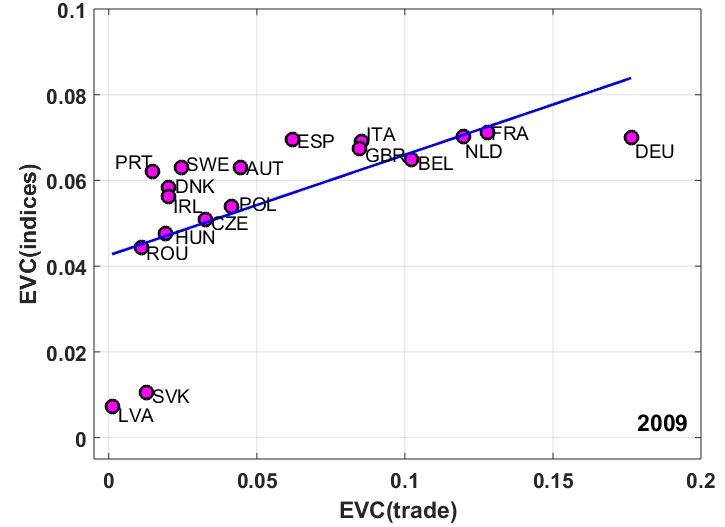}\llap{\parbox[b]{2in}{(h)\\\rule{0ex}{1.3in}}}\\
		\caption{Mapping between the EVC's of the financial network and the trade network for the years: 2003, 2005, 2007, 2009 across 18 European countries. The left panel (a,b,c and d) shows bar charts of the normalized EVC's of financial indices and trade. The right panel (e,f,g and h) shows the scattered plots of the normalized EVC's of financial indices and trade along with the best fit line having slope, (e) $0.43\pm0.09$ for 2003, (f) $0.40\pm0.09$ for 2005, (g) $0.25\pm0.08$ for 2007 and (h) $0.23\pm0.07$ for 2009. The positive slopes of the best fit line indicate that higher centrality in the financial network is correlated with occupying more central positions in the trade network. SVK, LVA and ROU always evolving together. DEU is an outlier.}
		\label{fig:Trade_Vs_Financial_Network}
	\end{figure}
	
	\begin{figure}[]
		\centering
		\includegraphics[width=0.42\linewidth]{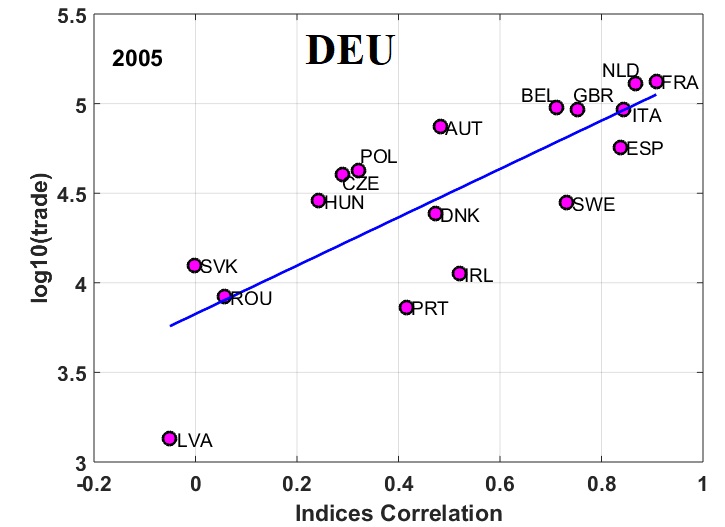}\llap{\parbox[b]{2.2in}{(a)\\\rule{0ex}{1.5in}}}
		\hspace{0.5em}
		\includegraphics[width=0.42\linewidth]{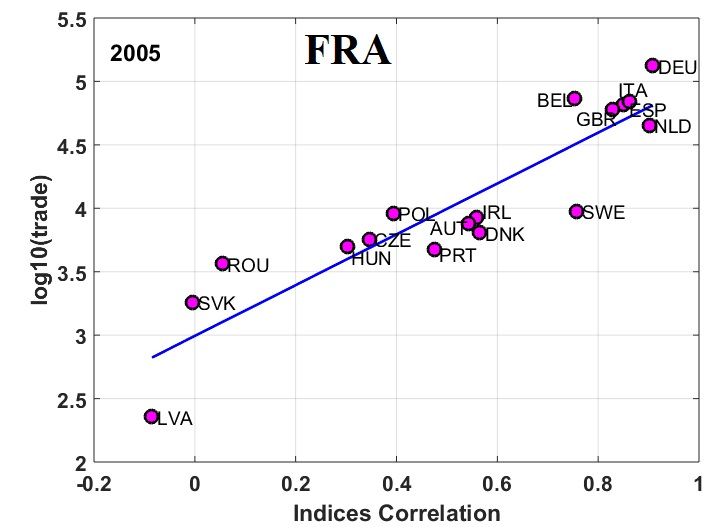}\llap{\parbox[b]{2.1in}{(e)\\\rule{0ex}{1.5in}}}
		\includegraphics[width=0.42\linewidth]{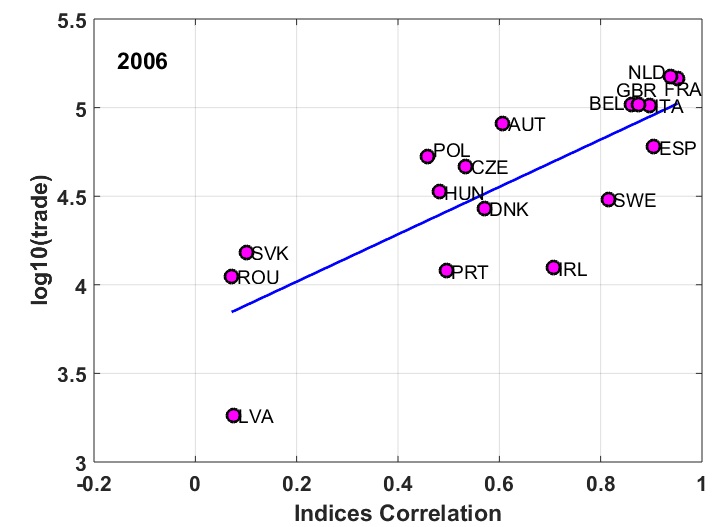}\llap{\parbox[b]{2.2in}{(b)\\\rule{0ex}{1.5in}}}
		\hspace{0.5em}
		\includegraphics[width=0.42\linewidth]{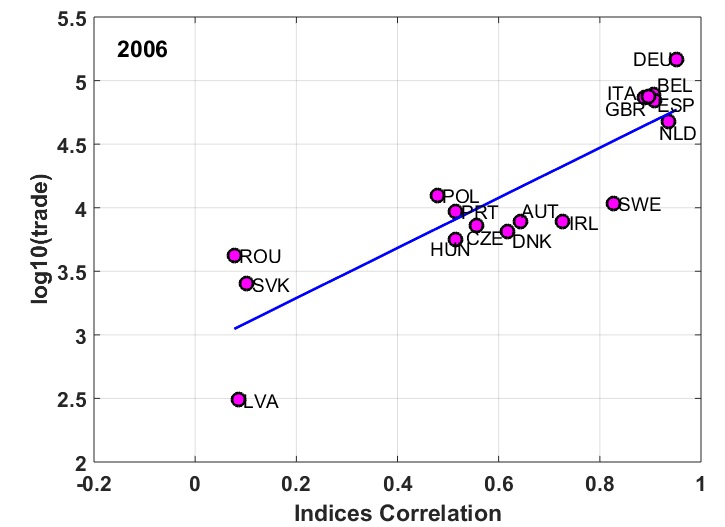}\llap{\parbox[b]{2.1in}{(f)\\\rule{0ex}{1.5in}}}
		\includegraphics[width=0.42\linewidth]{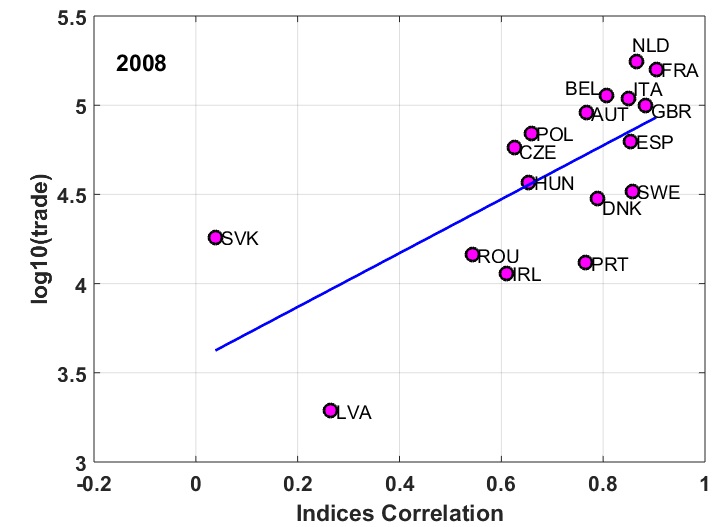}\llap{\parbox[b]{2.2in}{(c)\\\rule{0ex}{1.5in}}}
		\hspace{0.5em}
		\includegraphics[width=0.42\linewidth]{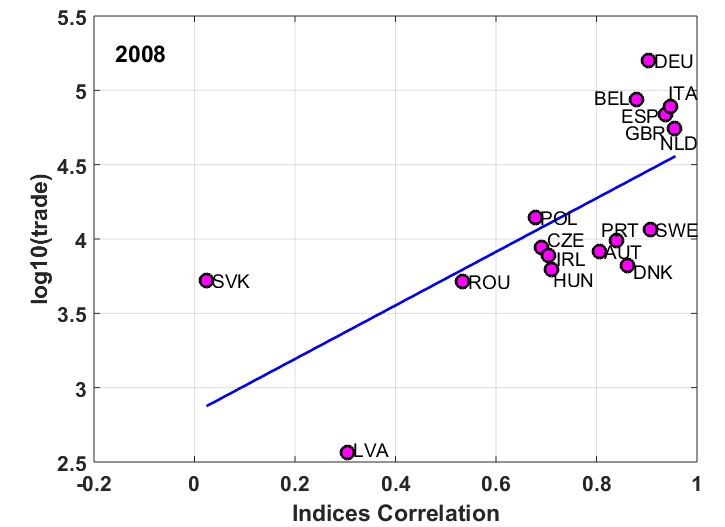}\llap{\parbox[b]{2.1in}{(g)\\\rule{0ex}{1.5in}}}
		\includegraphics[width=0.42\linewidth]{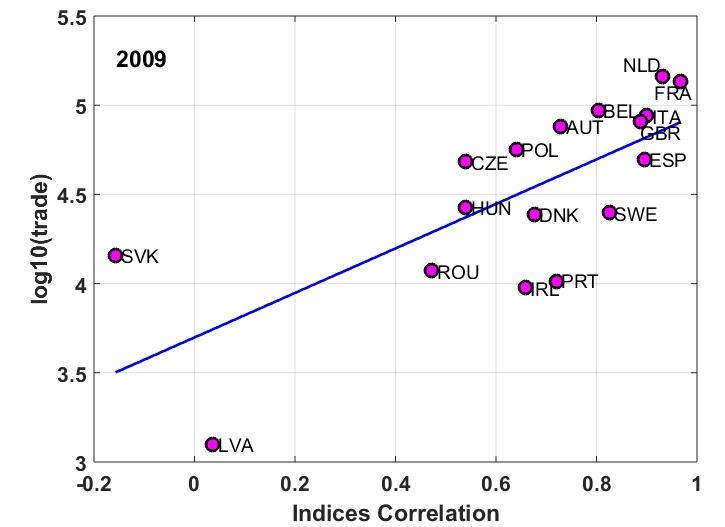}\llap{\parbox[b]{2.2in}{(d)\\\rule{0ex}{1.5in}}}
		\hspace{0.5em}
		\includegraphics[width=0.42\linewidth]{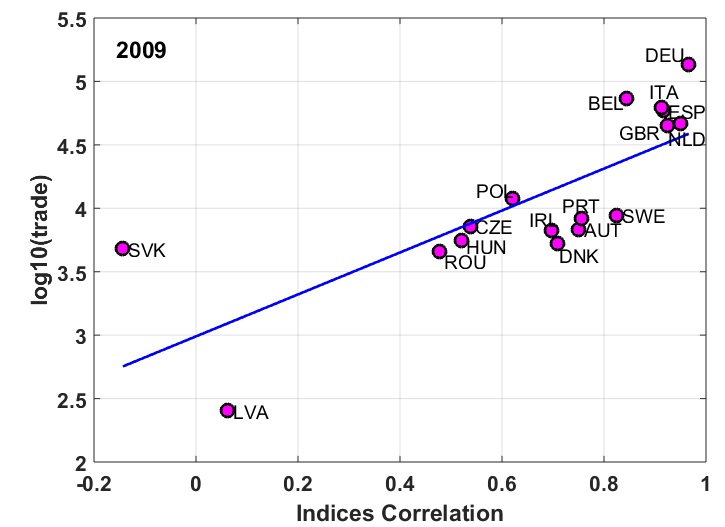}\llap{\parbox[b]{2.1in}{(h)\\\rule{0ex}{1.5in}}}
		
		\caption{Mapping between financial indices correlation and log10(trade) for Germany (DEU) and France (FRA), for four snapshots over time, two before the crisis (2005 and 2006) and two into the crisis period (2008 and 2009). For DEU the best fit line having slopes: (a) $1.35\pm0.27 $ for 2005, (b) $ 1.35\pm0.26 $ for 2006,(c) $1.51 \pm0.41 $ for 2008 and (d) $1.25 \pm 0.31$ for 2009. For FRA the best fit line having slope, (e) $2.00 \pm 0.23 $ for 2005, (f) $ 1.97 \pm 0.28$ for 2006, (g) $1.80\pm0.47 $ for 2008 and (h) $1.65 \pm0.35 $ for 2009.}
		\label{fig:DEU_FRA}
	\end{figure}
	
	We also conduct a microscopic study of the relation between trade flow and co-evolution of financial indices at the country level. For illustrative purpose, we have chosen two reasonably large European economies viz. Germany (DEU) and France (FRA). In Fig. \ref{fig:DEU_FRA}, we plot the nominal trade flow as a function of index correlations for pairs of countries, where we fix the origin country. In Fig. \ref{fig:DEU_FRA},the left column shows the analysis for Germany (DEU), whereas the right column shows the analysis for France (FRA). We have considered four snapshots over time, two before the crisis (2005 and 2006) and two into the crisis period (2008 and 2009).  For DEU the best fit line having slope: (a) $1.35\pm0.27 $ for 2005, (b) $ 1.35\pm0.26 $ for 2006,(c) $1.51 \pm0.41 $ for 2008, and (d) $1.25 \pm 0.31$ for 2009. For FRA the best fit line having slope: (e) $2.00 \pm 0.23 $ for 2005, (f) $ 1.97 \pm 0.28$ for 2006, (g) $1.80\pm0.47 $ for 2008, and (h) $1.65 \pm0.35 $ for 2009. One interesting feature is that during the crisis period, many countries become much more correlated and hence create a cluster, most notably in the case of Germany in periods 2008-09. However, in all the cases, it seems to be a clear positive correlation between pairwise trade flow and index correlation and this relationship is seemingly robust with respect to the occurrence of the crisis period. Three countries viz. Latvia (LVA), Romania (ROU) and Slovakia (SVK) seems to be far less correlated than the rest of the countries in the sample. However, removal of them does not affect the direction of the relationship.
	
	\subsection{ Mapping Between Economic Complexity Index (ECI) and Financial Indices}
	\label{subsub:ECI_equity}
	
	To find out the production characteristics of large economies, Economic Complexity Index (ECI) is a holistic measure proposed by Hidalgo and Hausmann \cite{Hidalgo_2009} in 2009. The goal of this index is to explain an economy as a whole rather than the sum of its parts. To see the mapping between equity and ECI, we regress the normalized EVC's of financial indices and ECI of 51 countries across the globe during 2007-2010, as shown in Fig. \ref{fig:EVC_Complexity}. Equity and ECI are sharing a positive relationship among themselves. Also the evolution of three variables: per capita GDP, ECI and EVC's of financial indices during 2002-2014 is shown in Fig. \ref{fig:evolution_ECI_EVC}.
	
	This finding is not very surprising as there are two fundamental relationships. One, typically larger (and more developed) countries have higher complexity index. Two, there is a strong relationship between return centrality and size (we explore it below in more details). Combining the two, we see that ECI could also have positive correlation with return centrality.

	\begin{figure}[]
		\centering
		\includegraphics[width=0.43\linewidth]{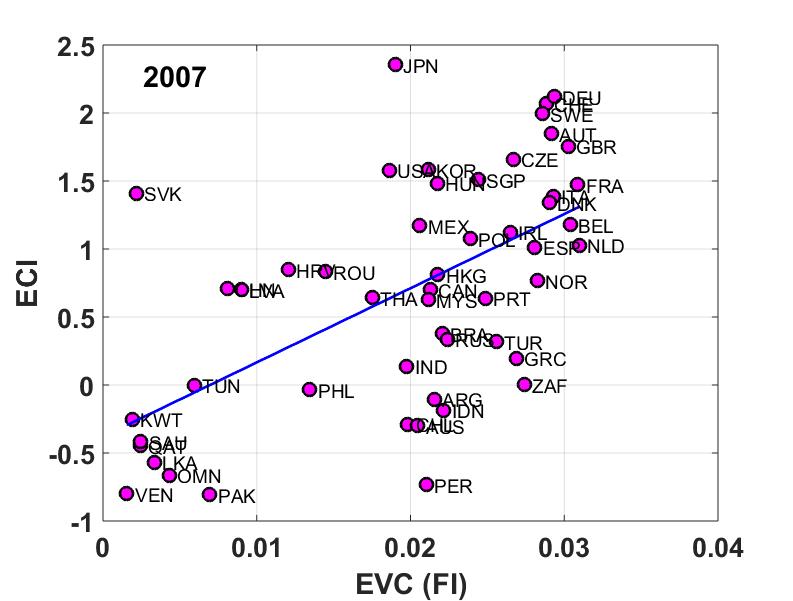}\llap{\parbox[b]{2.2in}{(a)\\\rule{0ex}{1.3in}}}
		\hspace{1em}
		\includegraphics[width=0.43\linewidth]{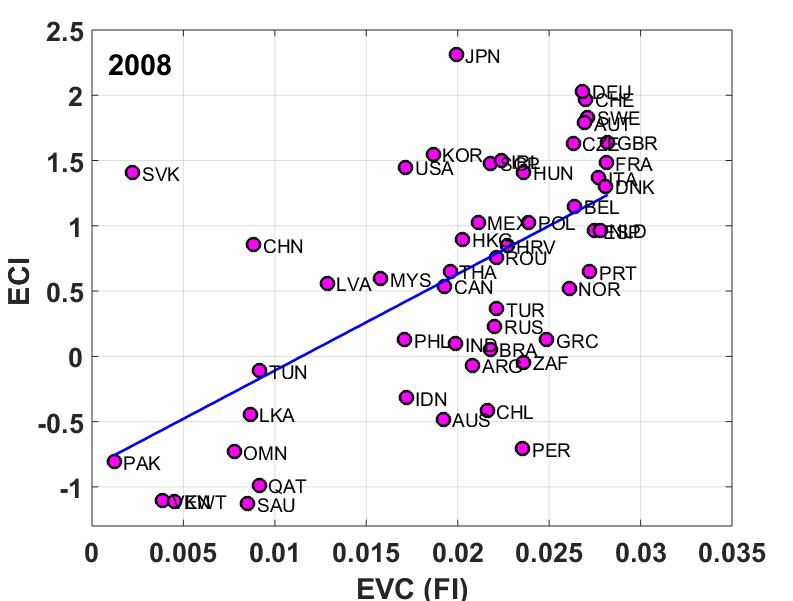}\llap{\parbox[b]{2.2in}{(b)\\\rule{0ex}{1.3in}}}
		\includegraphics[width=0.43\linewidth]{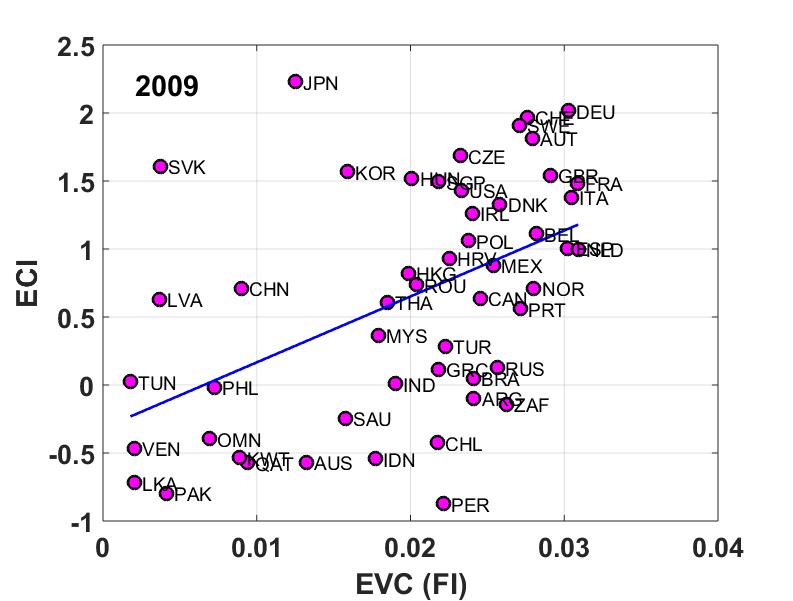}\llap{\parbox[b]{2.2in}{(c)\\\rule{0ex}{1.3in}}}
		\hspace{1em}
		\includegraphics[width=0.43\linewidth]{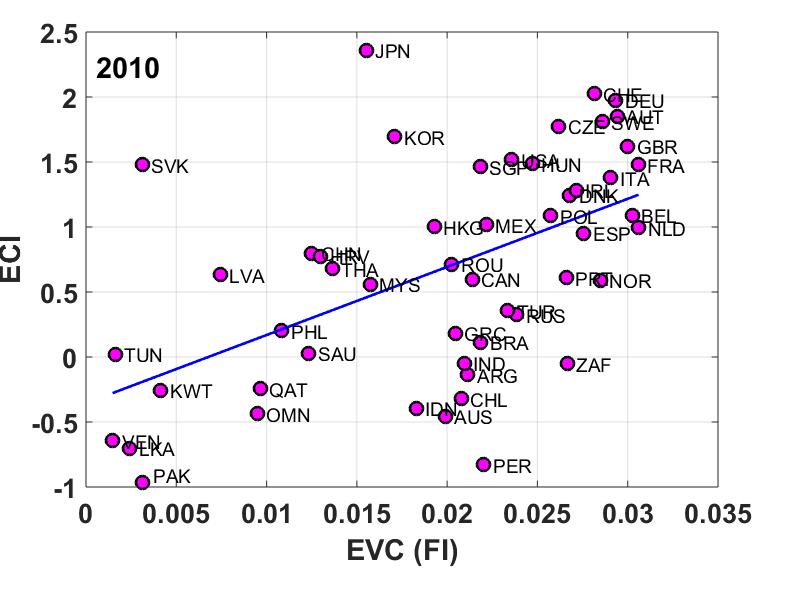}\llap{\parbox[b]{2.2in}{(d)\\\rule{0ex}{1.3in}}}
		\caption{Mapping between the EVC's of financial indices and economic complexity index (ECI), with the best fit lines having a slopes: (a) $ 54.5 \pm 10.7$ for 2007, (b) $ 73.9 \pm 13.9$ for 2008, (c) $ 48.5 \pm 12.3 $ for 2009, and (d) $ 52.4 \pm 11.7$ for 2010.  }
		\label{fig:EVC_Complexity}
	\end{figure}
	\begin{figure}[]
		\centering
		\includegraphics[width=0.85\linewidth]{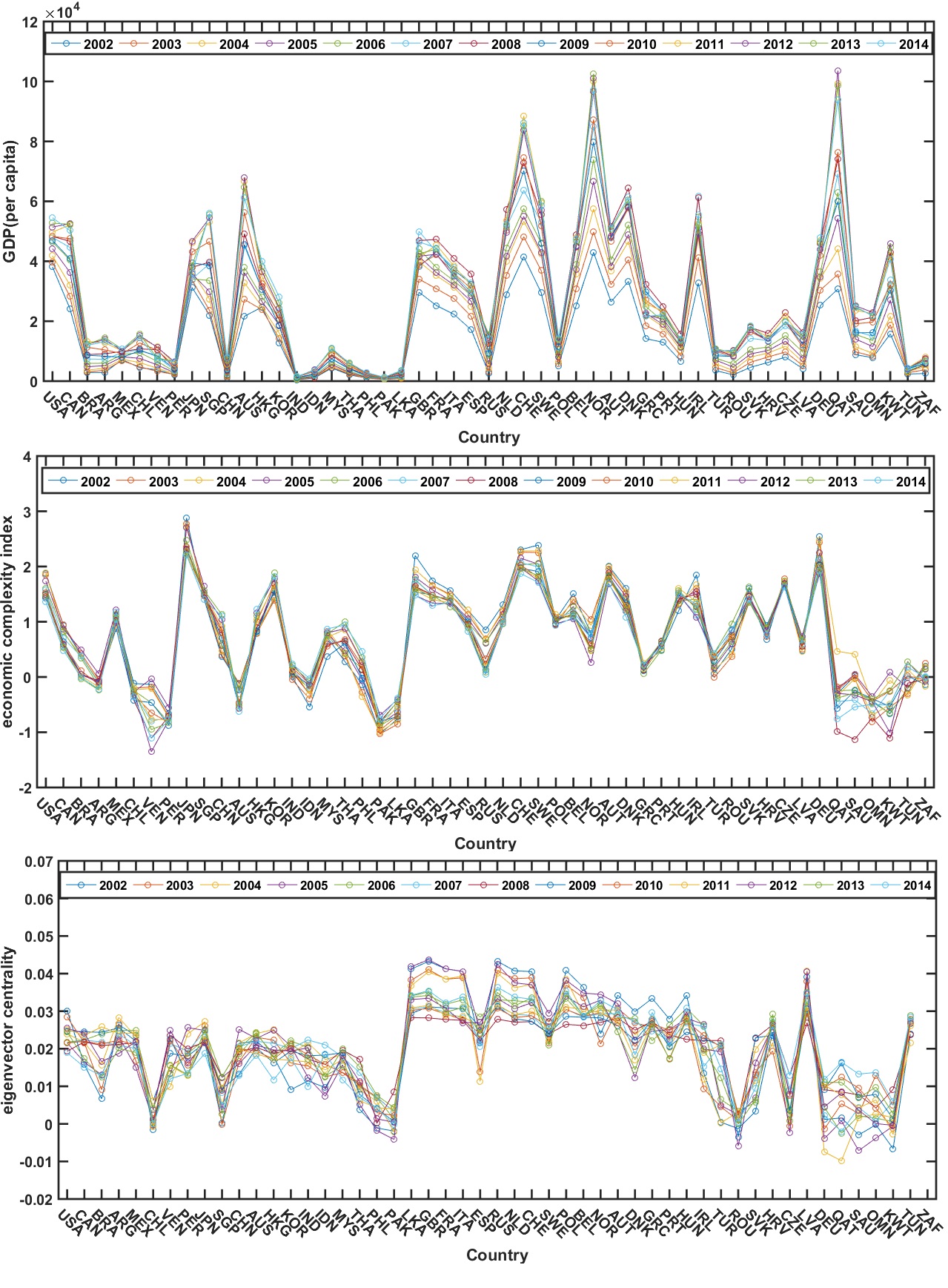}
		\llap{\parbox[b]{4.2in}{(a)\\\rule{0ex}{5.2in}}}
		\llap{\parbox[b]{4.2in}{(b)\\\rule{0ex}{3.4in}}}
		\llap{\parbox[b]{4.2in}{(c)\\\rule{0ex}{1.8in}}}
		\caption{Time evolution of (a) GDP per capita, (b) economic complexity index and (c) eigenvector centrality of market indices of different countries across the globe for the period 2002-2014.}
		\label{fig:evolution_ECI_EVC}
	\end{figure}

	\subsection{Estimation Results Controlling for Variations across Countries}

	All analyses done so far were essentially correlation study without controlling for other country-specific characteristics. Here, we present a sequence of regression tables done across years with control variables in place (2001-09; see Tables: 1 to 9 in arXiv:1805.06829). We have used foreign direct investment, total credit	as a percentage of GDP, trade openness (total trade/GDP), size variables (GDP and GDP per capita) as control variables. As can be seen, the relationship is not robust to inclusion of aggregate size (i.e. GDP). We have discussed this issue below in details.
   	
	Next, we have constructed an instrumental variable based on geographic centrality
	of the countries. The assumption we make is that geographic centrality should be orthogonal to size, but related to trade centrality (because of gravity equation; see below). The results are presented in Table 10 in arXiv:1805.06829 (2 stage least square estimation) and Table 11 in arXiv:1805.06829  (limited information maximum likelihood estimation). As can be seen, the sign is preserved and in the expected direction but the relationship is not statistically significant at 5\%. This is somewhat problematic as it indicates that the instrument is not very good for this test.
	
	Finally, we put all data into one balanced panel structure and find panel estimates without incorporating the control variables (unfortunately all data are not available). In this case, as can be expected, the relationship prevails (see Tables 12 and 13 in arXiv:1805.06829).
Hausman test confirms that a random effect model is more appropriate here. We checked if there is any relationship between EVC from trade and EVC from return across time (rather than across countries, as we have discussed above). In particular, it would be of interest to see if there is any strong indication of {\it Granger causality}. The results are presented in Tables 14 to 31 in arXiv:1805.06829. We see that there is no systematic relationship across these variables over time (we have included two lags for all estimations). Note that the time length is very small (9 years). Hence, we cannot infer much from the {\it VAR} analysis.

\subsection{Economic Interpretation and Econometric Issues} 

We have shown that there is a mapping between the networks of real and nominal variables. It is important to stress that this establishes the novelty of the present approach over and above the basic findings of Sharma et al. \cite{Sharma_2017_b}. The main statement of Sharma et al. \cite{Sharma_2017_b} is that centrality in the financial market is related to the size effect. In the present case, the same still holds true and that can be explained easily through gravity equation of trade, which states that the trade volume ($T_{ij}$) between two countries is approximately proportional to the product of the size of the countries ($Y_i$ and $Y_j$) and inversely proportional to their distance ($d_{ij}$). This can be stated as
	\begin{equation}
	T_{ij}\sim \frac{Y_i\times Y_j}{d_{ij}}.
	\end{equation}
	This implies that the EVC from the trade matrix is highly correlated with the size itself. This, in turn, implies that the relationship we find between the financial network and the trade network, may actually be a manifestation of the centrality-size relationship, similar to the finding in Sharma et al. \cite{Sharma_2017_b}.
	
	This raises a fundamental question about the nature of the relationship. Is it centrality-centrality or centrality-size? We cannot provide a complete answer to that. There are three points that need to be considered. First, centrality-size identification is an extremely difficult exercise as typically these two variables are highly correlated. Second, to characterize spill-over effects, network structures are useful whereas the size effect is not. Finally, the relationship between EVC of financial network and the trade network, is not monotonic. The linear fit captures the positive relationship. But a non-linear fit shows that the effect of higher centrality in trade diminishes after a steep initial increase. Thus, the multi-layered network view (with EVC-EVC as opposed to EVC-size relationship) is important to recognize the non-monotonic behavior.
	
\section{Snapshot of the World Stock Market}
	
	\begin{figure}[]
		\centering
		\includegraphics[width=0.85\linewidth]{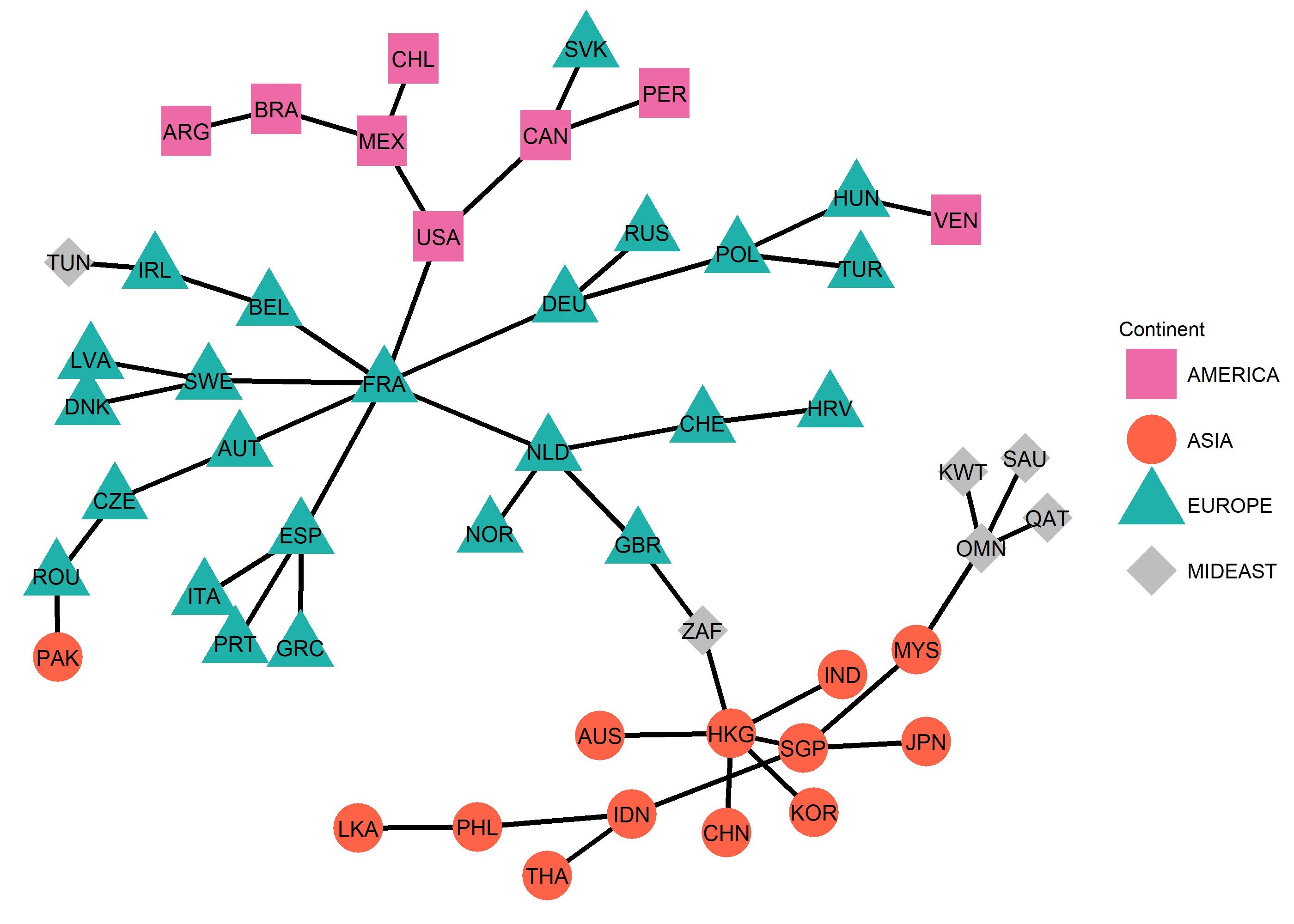}
		\caption{Minimum spanning tree of 51 market indices across the globe during the period 2013-2014. The MST shows $6$ African markets (gray diamonds), $13$ Asian markets (orange circles), $24$ European markets (green triangles) and $8$ Latin America markets (magenta squares).}
		\label{fig:world_MST}
	\end{figure}
	Fig. \ref{fig:world_MST} shows the minimum spanning tree (MST) of 51 market indices obtained from the Pearson cross-correlation matrix across the globe during the period 2013-2014. The nodes in the tree represent the market indices of the corresponding countries and the links between the nodes represent the relative distances of the distance matrix, $d=\sqrt{2(1-\rho)}$, where $\rho$ represents the correlation matrix. 
Thus, the minimum spanning tree reveals the structure of the global market indices and provides simple visualization about the patterns of links between different markets, similar to what was observed by Wang et al. \cite{Wang_2018}. 	
The MST indicates that geographic proximity plays big role in
	shaping up the correlation structure across markets.
	This feature has been noted and documented by other researchers as well \cite{Vodenska_2015}. One can conjecture that the main factor behind this
    observation is that financial markets react very quickly to news and hence,
    any bout of volatility in a market will be transmitted to another market when 
    that opens. For example, Tokyo stock exchange opens before London stock exchange. Hence, it is conceivable that there would be volatility spillover from Tokyo to 
    London. Although this qualitative explanation is intuitive, it remains unclear
    how to understand the underlying mechanism quantitatively.
    On a similar vein, it does not clearly explain the structure of the MST either.
    
\section{Summary}
    	
    In this chapter, we have demonstrated using multi-layered networks, the existence of an empirical linkage between the dynamics of the financial network constructed from the market indices and the macroeconomic networks constructed from macroeconomic variables such as trade, foreign direct investments, etc., for several countries across the globe. The time scales of the dynamics of the financial variables and the macroeconomic variables are orders of magnitude different, which makes the empirical linkage even more interesting and significant. Also, we found that there exist  in the respective networks, core-periphery structures (determined through eigenvector centrality measures) that are composed of similar sets of countries -- a result that may be related through the `gravity model' of the country-level macroeconomic networks. Thus, from a multi-lateral openness perspective, we showed that for individual countries, larger trade connectivity is positively associated with higher financial return correlations. We have specifically studied the two countries: Germany (DEU) and France (FRA), with respect to the other European countries. This revealed that mapping between the trade and financial indices correlation is quite robust across several years.
    
    Furthermore, we showed that the Economic Complexity Index and the equity markets have a positive relationship among themselves, as is the case for Gross Domestic Product; the time evolution of the three variables have interesting periodicities and correlation patterns. For certain countries the dispersions in the variables are rather pronounced than in other countries. To reveal the structure and dynamics of the global market indices, we have also studied the minimum spanning tree, which indicated that the geographical proximity does play an important role in the correlation structure across different markets. Perhaps the time-lagged correlation studies would reveal further the lead-lag structure of the markets.

    As noted by many researchers, network approach illuminates several interesting facets of the structure of global economy. However, standard econometric techniques show that all superficial observations are not necessarily robust. In particular, whenever one wants to move from correlations to causality, one has to use extra caution. In the end, we note a simple point. Many proposed empirical relationships in the econophysics literature fail the test for robustness (both in economic and statistical sense). Usage of econometrics combined with simple economic intuitions could remedy the problem to a large extent.

    \section*{Acknowledgement}
    ASC acknowledges the support by the institute grant (R\&P), IIM Ahmedabad. AC and KS acknowledge the support by grant number BT/BI/03/004/2003(C) of Govt. of India, Ministry of Science and Technology, Department of Biotechnology, Bioinformatics division, DST-PURSE of the Jawaharlal Nehru University, and University of Potential Excellence-II grant (Project ID-47) of the Jawaharlal Nehru University, New Delhi.	KS acknowledges the University Grants Commission (Ministry of Human Resource Development, Govt. of India) for her senior research fellowship.
    

    
    
    	
\begin{table}[h]
\caption{Estimates for the year 2001}
{
\def\sym#1{\ifmmode^{#1}\else\(^{#1}\)\fi}
\begin{tabular}{l*{6}{c}}
\hline\hline
            &\multicolumn{1}{c}{(1)}&\multicolumn{1}{c}{(2)}&\multicolumn{1}{c}{(3)}&\multicolumn{1}{c}{(4)}&\multicolumn{1}{c}{(5)}&\multicolumn{1}{c}{(6)}\\
            &\multicolumn{1}{c}{evc\_return}&\multicolumn{1}{c}{evc\_return}&\multicolumn{1}{c}{evc\_return}&\multicolumn{1}{c}{evc\_return}&\multicolumn{1}{c}{evc\_return}&\multicolumn{1}{c}{evc\_return}\\
\hline
evc\_trade   &       0.353\sym{***}&       0.353\sym{***}&       0.246\sym{*}  &       0.251\sym{*}  &      -0.108         &       0.156         \\
            &      (4.07)         &      (4.07)         &      (2.91)         &      (2.80)         &     (-0.91)         &      (1.67)         \\
[1em]
credit      &                     &                     &    0.000299\sym{*}  &    0.000301\sym{*}  &    0.000160         & -0.00000935         \\
            &                     &                     &      (2.64)         &      (2.57)         &      (1.71)         &     (-0.05)         \\
[1em]
tradeofgdp  &                     &                     &                     &   0.0000273         &    0.000211         &  -0.0000975         \\
            &                     &                     &                     &      (0.23)         &      (2.12)         &     (-0.80)         \\
\end{tabular}
}
\label{table:reg_01}
\end{table}

\begin{table}
\caption{Estimates for the year 2002}
{
\def\sym#1{\ifmmode^{#1}\else\(^{#1}\)\fi}
\begin{tabular}{l*{6}{c}}
\hline\hline
            &\multicolumn{1}{c}{(1)}&\multicolumn{1}{c}{(2)}&\multicolumn{1}{c}{(3)}&\multicolumn{1}{c}{(4)}&\multicolumn{1}{c}{(5)}&\multicolumn{1}{c}{(6)}\\
            &\multicolumn{1}{c}{evc\_return}&\multicolumn{1}{c}{evc\_return}&\multicolumn{1}{c}{evc\_return}&\multicolumn{1}{c}{evc\_return}&\multicolumn{1}{c}{evc\_return}&\multicolumn{1}{c}{evc\_return}\\
\hline
evc\_trade   &       0.366\sym{**} &       0.367\sym{**} &       0.262\sym{*}  &       0.245\sym{*}  &     -0.0773         &       0.135         \\
            &      (3.95)         &      (3.98)         &      (2.96)         &      (2.58)         &     (-0.52)         &      (1.45)         \\
[1em]
fdi         &                     &    0.000715         &    0.000593         &    0.000939         &    0.000101         &    0.000413         \\
            &                     &      (1.11)         &      (1.08)         &      (1.18)         &      (0.14)         &      (0.58)         \\
[1em]
credit      &                     &                     &    0.000283\sym{*}  &    0.000271\sym{*}  &    0.000177         &  -0.0000442         \\
            &                     &                     &      (2.56)         &      (2.35)         &      (1.72)         &     (-0.27)         \\
\end{tabular}
}
\label{table:reg_02}
\end{table}
===============================================
\begin{table}
\caption{Estimates for the year 2003}
{
\def\sym#1{\ifmmode^{#1}\else\(^{#1}\)\fi}
\begin{tabular}{l*{6}{c}}
\hline\hline
            &\multicolumn{1}{c}{(1)}&\multicolumn{1}{c}{(2)}&\multicolumn{1}{c}{(3)}&\multicolumn{1}{c}{(4)}&\multicolumn{1}{c}{(5)}&\multicolumn{1}{c}{(6)}\\
            &\multicolumn{1}{c}{evc\_return}&\multicolumn{1}{c}{evc\_return}&\multicolumn{1}{c}{evc\_return}&\multicolumn{1}{c}{evc\_return}&\multicolumn{1}{c}{evc\_return}&\multicolumn{1}{c}{evc\_return}\\
\hline
evc\_trade   &       0.429\sym{***}&       0.413\sym{***}&       0.319\sym{**} &       0.279\sym{**} &     -0.0563         &       0.151         \\
            &      (4.69)         &      (4.62)         &      (3.91)         &      (3.22)         &     (-0.51)         &      (2.11)         \\
[1em]
fdi         &                     &     0.00110         &    0.000848         &     0.00160         &     0.00131         &     0.00153\sym{*}  \\
            &                     &      (1.40)         &      (1.28)         &      (1.78)         &      (2.03)         &      (2.37)         \\
[1em]
credit      &                     &                     &    0.000287\sym{*}  &    0.000244\sym{*}  &    0.000152         &   -0.000110         \\
            &                     &                     &      (2.79)         &      (2.27)         &      (1.89)         &     (-0.88)         \\
\end{tabular}
}
\label{table:reg_03}
\end{table}

\begin{table}
\caption{Estimates for the year 2004}
{
\def\sym#1{\ifmmode^{#1}\else\(^{#1}\)\fi}
\begin{tabular}{l*{6}{c}}
\hline\hline
            &\multicolumn{1}{c}{(1)}&\multicolumn{1}{c}{(2)}&\multicolumn{1}{c}{(3)}&\multicolumn{1}{c}{(4)}&\multicolumn{1}{c}{(5)}&\multicolumn{1}{c}{(6)}\\
            &\multicolumn{1}{c}{evc\_return}&\multicolumn{1}{c}{evc\_return}&\multicolumn{1}{c}{evc\_return}&\multicolumn{1}{c}{evc\_return}&\multicolumn{1}{c}{evc\_return}&\multicolumn{1}{c}{evc\_return}\\
\hline
evc\_trade   &       0.422\sym{***}&       0.417\sym{***}&       0.318\sym{**} &       0.306\sym{**} &     -0.0316         &       0.135         \\
            &      (4.66)         &      (4.35)         &      (3.92)         &      (3.37)         &     (-0.25)         &      (1.86)         \\
[1em]
fdi         &                     &    0.000107         &    0.000196         &    0.000257         &    0.000249         &    0.000444         \\
            &                     &      (0.24)         &      (0.55)         &      (0.63)         &      (0.78)         &      (1.62)         \\
[1em]
credit      &                     &                     &    0.000311\sym{**} &    0.000304\sym{*}  &    0.000213\sym{*}  &  -0.0000848         \\
            &                     &                     &      (3.22)         &      (2.97)         &      (2.52)         &     (-0.74)         \\
\end{tabular}
}
\label{table:reg_04}
\end{table}

\begin{table}
\caption{Estimates for the year 2005}
{
\def\sym#1{\ifmmode^{#1}\else\(^{#1}\)\fi}
\begin{tabular}{l*{6}{c}}
\hline\hline
            &\multicolumn{1}{c}{(1)}&\multicolumn{1}{c}{(2)}&\multicolumn{1}{c}{(3)}&\multicolumn{1}{c}{(4)}&\multicolumn{1}{c}{(5)}&\multicolumn{1}{c}{(6)}\\
            &\multicolumn{1}{c}{evc\_return}&\multicolumn{1}{c}{evc\_return}&\multicolumn{1}{c}{evc\_return}&\multicolumn{1}{c}{evc\_return}&\multicolumn{1}{c}{evc\_return}&\multicolumn{1}{c}{evc\_return}\\
\hline
evc\_trade   &       0.399\sym{***}&       0.391\sym{***}&       0.337\sym{***}&       0.314\sym{**} &     -0.0591         &       0.168\sym{*}  \\
            &      (4.58)         &      (4.50)         &      (4.36)         &      (3.85)         &     (-0.49)         &      (2.19)         \\
[1em]
fdi         &                     &    0.000226         &   0.0000771         &    0.000220         &   0.0000846         &   0.0000973         \\
            &                     &      (1.09)         &      (0.41)         &      (0.92)         &      (0.48)         &      (0.52)         \\
[1em]
credit      &                     &                     &    0.000239\sym{*}  &    0.000194         &    0.000131         &  -0.0000941         \\
            &                     &                     &      (2.55)         &      (1.83)         &      (1.66)         &     (-0.79)         \\
\end{tabular}
}
\label{table:reg_05}
\end{table}

\begin{table}
\caption{Estimates for the year 2006}
{
\def\sym#1{\ifmmode^{#1}\else\(^{#1}\)\fi}
\begin{tabular}{l*{6}{c}}
\hline\hline
            &\multicolumn{1}{c}{(1)}&\multicolumn{1}{c}{(2)}&\multicolumn{1}{c}{(3)}&\multicolumn{1}{c}{(4)}&\multicolumn{1}{c}{(5)}&\multicolumn{1}{c}{(6)}\\
            &\multicolumn{1}{c}{evc\_return}&\multicolumn{1}{c}{evc\_return}&\multicolumn{1}{c}{evc\_return}&\multicolumn{1}{c}{evc\_return}&\multicolumn{1}{c}{evc\_return}&\multicolumn{1}{c}{evc\_return}\\
\hline
evc\_trade   &       0.327\sym{**} &       0.320\sym{**} &       0.286\sym{**} &       0.297\sym{**} &      -0.116         &       0.114         \\
            &      (3.79)         &      (3.48)         &      (3.42)         &      (3.10)         &     (-0.85)         &      (1.15)         \\
[1em]
fdi         &                     &   0.0000490         &   0.0000300         &  0.00000673         &   0.0000431         &   0.0000837         \\
            &                     &      (0.29)         &      (0.20)         &      (0.04)         &      (0.33)         &      (0.57)         \\
[1em]
credit      &                     &                     &    0.000187\sym{*}  &    0.000195         &    0.000110         &  -0.0000853         \\
            &                     &                     &      (2.16)         &      (2.06)         &      (1.49)         &     (-0.69)         \\
\end{tabular}
}
\label{table:reg_06}
\end{table}

\begin{table}
\caption{Estimates for the year 2007}
{
\def\sym#1{\ifmmode^{#1}\else\(^{#1}\)\fi}
\begin{tabular}{l*{6}{c}}
\hline\hline
            &\multicolumn{1}{c}{(1)}&\multicolumn{1}{c}{(2)}&\multicolumn{1}{c}{(3)}&\multicolumn{1}{c}{(4)}&\multicolumn{1}{c}{(5)}&\multicolumn{1}{c}{(6)}\\
            &\multicolumn{1}{c}{evc\_return}&\multicolumn{1}{c}{evc\_return}&\multicolumn{1}{c}{evc\_return}&\multicolumn{1}{c}{evc\_return}&\multicolumn{1}{c}{evc\_return}&\multicolumn{1}{c}{evc\_return}\\
\hline
evc\_trade   &       0.248\sym{**} &       0.240\sym{*}  &       0.219\sym{**} &       0.198\sym{*}  &     -0.0953         &      0.0525         \\
            &      (3.12)         &      (2.90)         &      (2.99)         &      (2.47)         &     (-0.76)         &      (0.59)         \\
[1em]
fdi         &                     &   0.0000494         &   0.0000474         &   0.0000923         &   0.0000965         &    0.000124         \\
            &                     &      (0.49)         &      (0.53)         &      (0.84)         &      (1.07)         &      (1.32)         \\
[1em]
credit      &                     &                     &    0.000166\sym{*}  &    0.000147         &   0.0000819         &  -0.0000539         \\
            &                     &                     &      (2.35)         &      (1.92)         &      (1.23)         &     (-0.53)         \\
\end{tabular}
}
\label{table:reg_07}
\end{table}

\begin{table}
\caption{Estimates for the year 2008}
{
\def\sym#1{\ifmmode^{#1}\else\(^{#1}\)\fi}
\begin{tabular}{l*{6}{c}}
\hline\hline
            &\multicolumn{1}{c}{(1)}&\multicolumn{1}{c}{(2)}&\multicolumn{1}{c}{(3)}&\multicolumn{1}{c}{(4)}&\multicolumn{1}{c}{(5)}&\multicolumn{1}{c}{(6)}\\
            &\multicolumn{1}{c}{evc\_return}&\multicolumn{1}{c}{evc\_return}&\multicolumn{1}{c}{evc\_return}&\multicolumn{1}{c}{evc\_return}&\multicolumn{1}{c}{evc\_return}&\multicolumn{1}{c}{evc\_return}\\
\hline
evc\_trade   &       0.167\sym{*}  &       0.164\sym{*}  &       0.154\sym{*}  &       0.104         &     -0.0813         &     -0.0359         \\
            &      (2.30)         &      (2.18)         &      (2.30)         &      (1.60)         &     (-0.68)         &     (-0.46)         \\
[1em]
fdi         &                     &   0.0000510         &   0.0000871         &    0.000294         &    0.000273         &    0.000385\sym{*}  \\
            &                     &      (0.38)         &      (0.72)         &      (2.01)         &      (2.01)         &      (2.99)         \\
[1em]
credit      &                     &                     &    0.000141\sym{*}  &    0.000112         &   0.0000620         &  -0.0000414         \\
            &                     &                     &      (2.22)         &      (1.92)         &      (1.02)         &     (-0.53)         \\
\end{tabular}
}
\label{table:reg_08}
\end{table}

\begin{table}
\caption{Estimates for the year 2009}
{
\def\sym#1{\ifmmode^{#1}\else\(^{#1}\)\fi}
\begin{tabular}{l*{6}{c}}
\hline\hline
            &\multicolumn{1}{c}{(1)}&\multicolumn{1}{c}{(2)}&\multicolumn{1}{c}{(3)}&\multicolumn{1}{c}{(4)}&\multicolumn{1}{c}{(5)}&\multicolumn{1}{c}{(6)}\\
            &\multicolumn{1}{c}{evc\_return}&\multicolumn{1}{c}{evc\_return}&\multicolumn{1}{c}{evc\_return}&\multicolumn{1}{c}{evc\_return}&\multicolumn{1}{c}{evc\_return}&\multicolumn{1}{c}{evc\_return}\\
\hline
evc\_trade   &       0.235\sym{**} &       0.229\sym{**} &       0.229\sym{**} &       0.167\sym{*}  &      -0.124         &       0.100         \\
            &      (3.16)         &      (3.01)         &      (3.01)         &      (2.17)         &     (-1.38)         &      (1.31)         \\
[1em]
fdi         &                     &    0.000193         &    0.000193         &    0.000675         &    0.000342         &    0.000305         \\
            &                     &      (0.63)         &      (0.63)         &      (1.78)         &      (1.24)         &      (0.79)         \\
[1em]
tradeofgdp  &                     &                     &                     &   -0.000240         &   0.0000132         &   -0.000194         \\
            &                     &                     &                     &     (-1.91)         &      (0.12)         &     (-1.68)         \\
\end{tabular}
}
\label{table:reg_09}
\end{table}

===============================================
\begin{sidewaystable} 
\centering
\caption{All IV estimation (2SLS)}
{
\def\sym#1{\ifmmode^{#1}\else\(^{#1}\)\fi}
\begin{tabular}{l*{9}{c}}
\hline\hline
            &\multicolumn{1}{c}{(1)}&\multicolumn{1}{c}{(2)}&\multicolumn{1}{c}{(3)}&\multicolumn{1}{c}{(4)}&\multicolumn{1}{c}{(5)}&\multicolumn{1}{c}{(6)}&\multicolumn{1}{c}{(7)}&\multicolumn{1}{c}{(8)}&\multicolumn{1}{c}{(9)}\\
            &\multicolumn{1}{c}{evc\_return}&\multicolumn{1}{c}{evc\_return}&\multicolumn{1}{c}{evc\_return}&\multicolumn{1}{c}{evc\_return}&\multicolumn{1}{c}{evc\_return}&\multicolumn{1}{c}{evc\_return}&\multicolumn{1}{c}{evc\_return}&\multicolumn{1}{c}{evc\_return}&\multicolumn{1}{c}{evc\_return}\\
\hline
evc\_trade   &      0.0318         &       0.108         &       0.121         &       0.134         &       0.153         &       0.147         &       0.120         &      0.0337         &      0.0965         \\
            &      (0.15)         &      (0.69)         &      (0.88)         &      (0.84)         &      (0.91)         &      (0.78)         &      (0.74)         &      (0.26)         &      (0.73)         \\
[1em]
credit      &    0.000430\sym{**} &    0.000334\sym{**} &    0.000290\sym{**} &    0.000354\sym{***}&    0.000213\sym{*}  &    0.000197\sym{*}  &    0.000145\sym{*}  &    0.000110\sym{*}  &                     \\
            &      (2.62)         &      (2.76)         &      (2.71)         &      (3.35)         &      (2.05)         &      (2.25)         &      (2.15)         &      (2.13)         &                     \\
[1em]
tradeofgdp  &  -0.0000370         &   -0.000193         &   -0.000340         &   -0.000151         &   -0.000230         &  -0.0000567         &   -0.000128         &   -0.000266\sym{*}  &   -0.000289\sym{*}  \\
            &     (-0.27)         &     (-1.04)         &     (-1.78)         &     (-0.95)         &     (-1.40)         &     (-0.35)         &     (-0.97)         &     (-2.26)         &     (-2.10)         \\
\end{tabular}
}
\label{table:IVreg_1}
\end{sidewaystable}


\begin{sidewaystable} 
\centering
\caption{All IV estimation (LIML)}
{
\def\sym#1{\ifmmode^{#1}\else\(^{#1}\)\fi}
\begin{tabular}{l*{9}{c}}
\hline\hline
            &\multicolumn{1}{c}{(1)}&\multicolumn{1}{c}{(2)}&\multicolumn{1}{c}{(3)}&\multicolumn{1}{c}{(4)}&\multicolumn{1}{c}{(5)}&\multicolumn{1}{c}{(6)}&\multicolumn{1}{c}{(7)}&\multicolumn{1}{c}{(8)}&\multicolumn{1}{c}{(9)}\\
            &\multicolumn{1}{c}{evc\_return}&\multicolumn{1}{c}{evc\_return}&\multicolumn{1}{c}{evc\_return}&\multicolumn{1}{c}{evc\_return}&\multicolumn{1}{c}{evc\_return}&\multicolumn{1}{c}{evc\_return}&\multicolumn{1}{c}{evc\_return}&\multicolumn{1}{c}{evc\_return}&\multicolumn{1}{c}{evc\_return}\\
\hline
evc\_trade   &      0.0318         &       0.108         &       0.121         &       0.134         &       0.153         &       0.147         &       0.120         &      0.0337         &      0.0965         \\
            &      (0.15)         &      (0.69)         &      (0.88)         &      (0.84)         &      (0.91)         &      (0.78)         &      (0.74)         &      (0.26)         &      (0.73)         \\
[1em]
credit      &    0.000430\sym{**} &    0.000334\sym{**} &    0.000290\sym{**} &    0.000354\sym{***}&    0.000213\sym{*}  &    0.000197\sym{*}  &    0.000145\sym{*}  &    0.000110\sym{*}  &                     \\
            &      (2.62)         &      (2.76)         &      (2.71)         &      (3.35)         &      (2.05)         &      (2.25)         &      (2.15)         &      (2.13)         &                     \\
[1em]
tradeofgdp  &  -0.0000370         &   -0.000193         &   -0.000340         &   -0.000151         &   -0.000230         &  -0.0000567         &   -0.000128         &   -0.000266\sym{*}  &   -0.000289\sym{*}  \\
            &     (-0.27)         &     (-1.04)         &     (-1.78)         &     (-0.95)         &     (-1.40)         &     (-0.35)         &     (-0.97)         &     (-2.26)         &     (-2.10)         \\
\end{tabular}
}
\label{table:IVreg_2}
\end{sidewaystable}

===============================================
\begin{table}
\caption{Panel estimates: fixed and random effects}
\centering
{
\def\sym#1{\ifmmode^{#1}\else\(^{#1}\)\fi}
\begin{tabular}{l*{2}{c}}
\hline\hline
            &\multicolumn{1}{c}{(1)}&\multicolumn{1}{c}{(2)}\\
            &\multicolumn{1}{c}{evc\_return}&\multicolumn{1}{c}{evc\_return}\\
\hline
evc\_trade   &       0.550\sym{***}&       0.391\sym{***}\\
            &      (4.23)         &      (5.64)         \\
[1em]
\_cons      &      0.0250\sym{***}&      0.0339\sym{***}\\
            &      (3.45)         &      (6.12)         \\
\hline
\(N\)       &         162         &         162         \\
adj. \(R^{2}\)&      -0.001         &                     \\
\hline\hline
\multicolumn{3}{l}{\footnotesize \textit{t} statistics in parentheses}\\
\multicolumn{3}{l}{\footnotesize \sym{*} \(p<0.05\), \sym{**} \(p<0.01\), \sym{***} \(p<0.001\)}\\
\end{tabular}
}
\label{table:panel_reg_1}
\end{table}

\begin{table}
	\caption{Hausman test: Random effect is more appropriate.}
\centering
{
\def\sym#1{\ifmmode^{#1}\else\(^{#1}\)\fi}
\begin{tabular}{l*{1}{c}}
\hline\hline
            &\multicolumn{1}{c}{(1)}\\
            &\multicolumn{1}{c}{evc\_return}\\
\hline
evc\_trade   &       0.391\sym{***}\\
            &      (5.64)         \\
[1em]
\_cons      &      0.0339\sym{***}\\
            &      (6.12)         \\
\hline
\(N\)       &         162         \\
adj. \(R^{2}\)&                     \\
\hline\hline
\multicolumn{2}{l}{\footnotesize \textit{t} statistics in parentheses}\\
\multicolumn{2}{l}{\footnotesize \sym{*} \(p<0.05\), \sym{**} \(p<0.01\), \sym{***} \(p<0.001\)}\\
\end{tabular}
}
\label{table:panel_reg_2}
\end{table}
===============================================

\begin{table}
\caption{VAR estimation: Belgium (BEL)}
\centering
{
\def\sym#1{\ifmmode^{#1}\else\(^{#1}\)\fi}
\begin{tabular}{l*{1}{c}}
\hline\hline
            &\multicolumn{1}{c}{(1)}\\
            &\multicolumn{1}{c}{bel\_r}\\
\hline
bel\_r       &                     \\
L.bel\_r     &      -0.412         \\
            &     (-1.05)         \\
[1em]
L2.bel\_r    &       0.548         \\
            &      (1.68)         \\
[1em]
L.bel\_t     &      -1.901\sym{***}\\
            &     (-3.58)         \\
[1em]
L2.bel\_t    &      -1.042\sym{**} \\
            &     (-2.66)         \\
[1em]
\_cons      &       0.357\sym{***}\\
            &      (4.37)         \\
\hline
bel\_t       &                     \\
L.bel\_r     &       0.572\sym{**} \\
            &      (3.14)         \\
[1em]
L2.bel\_r    &      -0.298\sym{*}  \\
            &     (-1.98)         \\
[1em]
L.bel\_t     &       0.425         \\
            &      (1.73)         \\
[1em]
L2.bel\_t    &       0.948\sym{***}\\
            &      (5.22)         \\
[1em]
\_cons      &     -0.0557         \\
            &     (-1.47)         \\
\hline
\(N\)       &           7         \\
\hline\hline
\multicolumn{2}{l}{\footnotesize \textit{t} statistics in parentheses}\\
\multicolumn{2}{l}{\footnotesize \sym{*} \(p<0.05\), \sym{**} \(p<0.01\), \sym{***} \(p<0.001\)}\\
\end{tabular}
}
\label{var:bel}
\end{table}
===============================================
\begin{table}
	\label{var:cze}
\caption{VAR estimation: Czech Republic (CZE)}
\centering
{
\def\sym#1{\ifmmode^{#1}\else\(^{#1}\)\fi}
\begin{tabular}{l*{1}{c}}
\hline\hline
            &\multicolumn{1}{c}{(1)}\\
            &\multicolumn{1}{c}{cze\_r}\\
\hline
cze\_r       &                     \\
L.cze\_r     &       0.450         \\
            &      (1.24)         \\
[1em]
L2.cze\_r    &      -0.705\sym{*}  \\
            &     (-2.22)         \\
[1em]
L.cze\_t     &       2.064         \\
            &      (1.19)         \\
[1em]
L2.cze\_t    &      -0.433         \\
            &     (-0.25)         \\
[1em]
\_cons      &      0.0162         \\
            &      (1.49)         \\
\hline
cze\_t       &                     \\
L.cze\_r     &       0.133         \\
            &      (1.76)         \\
[1em]
L2.cze\_r    &     -0.0718         \\
            &     (-1.09)         \\
[1em]
L.cze\_t     &       0.790\sym{*}  \\
            &      (2.19)         \\
[1em]
L2.cze\_t    &    -0.00296         \\
            &     (-0.01)         \\
[1em]
\_cons      &     0.00421         \\
            &      (1.87)         \\
\hline
\(N\)       &           7         \\
\hline\hline
\multicolumn{2}{l}{\footnotesize \textit{t} statistics in parentheses}\\
\multicolumn{2}{l}{\footnotesize \sym{*} \(p<0.05\), \sym{**} \(p<0.01\), \sym{***} \(p<0.001\)}\\
\end{tabular}
}
\end{table}
===============================================
\begin{table}
	\label{var:dnk}
\caption{VAR estimation: Denmark (DNK)}
\centering
{
\def\sym#1{\ifmmode^{#1}\else\(^{#1}\)\fi}
\begin{tabular}{l*{1}{c}}
\hline\hline
            &\multicolumn{1}{c}{(1)}\\
            &\multicolumn{1}{c}{dnk\_r}\\
\hline
dnk\_r       &                     \\
L.dnk\_r     &       0.122         \\
            &      (0.67)         \\
[1em]
L2.dnk\_r    &      -0.950\sym{***}\\
            &     (-4.77)         \\
[1em]
L.dnk\_t     &       1.195         \\
            &      (1.70)         \\
[1em]
L2.dnk\_t    &       2.277\sym{**} \\
            &      (2.87)         \\
[1em]
\_cons      &      0.0388         \\
            &      (1.69)         \\
\hline
dnk\_t       &                     \\
L.dnk\_r     &     -0.0774\sym{*}  \\
            &     (-1.98)         \\
[1em]
L2.dnk\_r    &      0.0317         \\
            &      (0.74)         \\
[1em]
L.dnk\_t     &       0.195         \\
            &      (1.29)         \\
[1em]
L2.dnk\_t    &      -0.240         \\
            &     (-1.40)         \\
[1em]
\_cons      &      0.0243\sym{***}\\
            &      (4.89)         \\
\hline
\(N\)       &           7         \\
\hline\hline
\multicolumn{2}{l}{\footnotesize \textit{t} statistics in parentheses}\\
\multicolumn{2}{l}{\footnotesize \sym{*} \(p<0.05\), \sym{**} \(p<0.01\), \sym{***} \(p<0.001\)}\\
\end{tabular}
}
\end{table}
===============================================
\begin{table}
	\label{var:deu}
\caption{VAR estimation: Germany (DEU)}
\centering
{
\def\sym#1{\ifmmode^{#1}\else\(^{#1}\)\fi}
\begin{tabular}{l*{1}{c}}
\hline\hline
            &\multicolumn{1}{c}{(1)}\\
            &\multicolumn{1}{c}{deu\_r}\\
\hline
deu\_r       &                     \\
L.deu\_r     &       1.396\sym{***}\\
            &      (3.60)         \\
[1em]
L2.deu\_r    &      -1.618\sym{**} \\
            &     (-2.60)         \\
[1em]
L.deu\_t     &      -1.992         \\
            &     (-1.55)         \\
[1em]
L2.deu\_t    &       2.733         \\
            &      (1.12)         \\
[1em]
\_cons      &     -0.0339         \\
            &     (-0.08)         \\
\hline
deu\_t       &                     \\
L.deu\_r     &     -0.0789\sym{**} \\
            &     (-3.20)         \\
[1em]
L2.deu\_r    &      -0.137\sym{***}\\
            &     (-3.48)         \\
[1em]
L.deu\_t     &       0.509\sym{***}\\
            &      (6.24)         \\
[1em]
L2.deu\_t    &      0.0607         \\
            &      (0.39)         \\
[1em]
\_cons      &      0.0913\sym{***}\\
            &      (3.40)         \\
\hline
\(N\)       &           7         \\
\hline\hline
\multicolumn{2}{l}{\footnotesize \textit{t} statistics in parentheses}\\
\multicolumn{2}{l}{\footnotesize \sym{*} \(p<0.05\), \sym{**} \(p<0.01\), \sym{***} \(p<0.001\)}\\
\end{tabular}
}
\end{table}

\begin{table}
	\label{var:ire}
\caption{VAR estimation: Ireland (IRE)}
\centering
{
\def\sym#1{\ifmmode^{#1}\else\(^{#1}\)\fi}
\begin{tabular}{l*{1}{c}}
\hline\hline
            &\multicolumn{1}{c}{(1)}\\
            &\multicolumn{1}{c}{deu\_r}\\
\hline
deu\_r       &                     \\
L.deu\_r     &       1.396\sym{***}\\
            &      (3.60)         \\
[1em]
L2.deu\_r    &      -1.618\sym{**} \\
            &     (-2.60)         \\
[1em]
L.deu\_t     &      -1.992         \\
            &     (-1.55)         \\
[1em]
L2.deu\_t    &       2.733         \\
            &      (1.12)         \\
[1em]
\_cons      &     -0.0339         \\
            &     (-0.08)         \\
\hline
deu\_t       &                     \\
L.deu\_r     &     -0.0789\sym{**} \\
            &     (-3.20)         \\
[1em]
L2.deu\_r    &      -0.137\sym{***}\\
            &     (-3.48)         \\
[1em]
L.deu\_t     &       0.509\sym{***}\\
            &      (6.24)         \\
[1em]
L2.deu\_t    &      0.0607         \\
            &      (0.39)         \\
[1em]
\_cons      &      0.0913\sym{***}\\
            &      (3.40)         \\
\hline
\(N\)       &           7         \\
\hline\hline
\multicolumn{2}{l}{\footnotesize \textit{t} statistics in parentheses}\\
\multicolumn{2}{l}{\footnotesize \sym{*} \(p<0.05\), \sym{**} \(p<0.01\), \sym{***} \(p<0.001\)}\\
\end{tabular}
}
\end{table}

\newpage

\begin{table}
	\label{var:esp}
\caption{VAR estimation: Spain (ESP)}
\centering
{
\def\sym#1{\ifmmode^{#1}\else\(^{#1}\)\fi}
\begin{tabular}{l*{1}{c}}
\hline\hline
            &\multicolumn{1}{c}{(1)}\\
            &\multicolumn{1}{c}{esp\_r}\\
\hline
esp\_r       &                     \\
L.esp\_r     &       1.259\sym{*}  \\
            &      (2.11)         \\
[1em]
L2.esp\_r    &       0.769         \\
            &      (0.85)         \\
[1em]
L.esp\_t     &      -1.501         \\
            &     (-0.85)         \\
[1em]
L2.esp\_t    &      -2.075         \\
            &     (-1.37)         \\
[1em]
\_cons      &       0.166         \\
            &      (1.78)         \\
\hline
esp\_t       &                     \\
L.esp\_r     &       0.755\sym{***}\\
            &      (5.90)         \\
[1em]
L2.esp\_r    &       0.789\sym{***}\\
            &      (4.07)         \\
[1em]
L.esp\_t     &      -1.124\sym{**} \\
            &     (-2.96)         \\
[1em]
L2.esp\_t    &      -1.201\sym{***}\\
            &     (-3.71)         \\
[1em]
\_cons      &       0.110\sym{***}\\
            &      (5.50)         \\
\hline
\(N\)       &           7         \\
\hline\hline
\multicolumn{2}{l}{\footnotesize \textit{t} statistics in parentheses}\\
\multicolumn{2}{l}{\footnotesize \sym{*} \(p<0.05\), \sym{**} \(p<0.01\), \sym{***} \(p<0.001\)}\\
\end{tabular}
}
\end{table}

\begin{table}
	\label{var:fra}
\caption{VAR estimation: France (FRA)}
\centering
{
\def\sym#1{\ifmmode^{#1}\else\(^{#1}\)\fi}
\begin{tabular}{l*{1}{c}}
\hline\hline
            &\multicolumn{1}{c}{(1)}\\
            &\multicolumn{1}{c}{fra\_r}\\
\hline
fra\_r       &                     \\
L.fra\_r     &      -0.437         \\
            &     (-1.27)         \\
[1em]
L2.fra\_r    &      0.0680         \\
            &      (0.19)         \\
[1em]
L.fra\_t     &       2.239\sym{*}  \\
            &      (2.15)         \\
[1em]
L2.fra\_t    &      -0.303         \\
            &     (-0.28)         \\
[1em]
\_cons      &      -0.147\sym{***}\\
            &     (-4.19)         \\
\hline
fra\_t       &                     \\
L.fra\_r     &      -0.259         \\
            &     (-1.55)         \\
[1em]
L2.fra\_r    &      -0.314         \\
            &     (-1.78)         \\
[1em]
L.fra\_t     &       1.331\sym{**} \\
            &      (2.64)         \\
[1em]
L2.fra\_t    &     0.00892         \\
            &      (0.02)         \\
[1em]
\_cons      &  -0.0000508         \\
            &     (-0.00)         \\
\hline
\(N\)       &           7         \\
\hline\hline
\multicolumn{2}{l}{\footnotesize \textit{t} statistics in parentheses}\\
\multicolumn{2}{l}{\footnotesize \sym{*} \(p<0.05\), \sym{**} \(p<0.01\), \sym{***} \(p<0.001\)}\\
\end{tabular}
}
\end{table}

\begin{table}
\label{var:ita}
\caption{VAR estimation: Italy (Ita)}
\centering
{
\def\sym#1{\ifmmode^{#1}\else\(^{#1}\)\fi}
\begin{tabular}{l*{1}{c}}
\hline\hline
            &\multicolumn{1}{c}{(1)}\\
            &\multicolumn{1}{c}{ita\_r}\\
\hline
ita\_r       &                     \\
L.ita\_r     &       0.320         \\
            &      (0.64)         \\
[1em]
L2.ita\_r    &      -1.037         \\
            &     (-1.64)         \\
[1em]
L.ita\_t     &       0.625\sym{*}  \\
            &      (2.23)         \\
[1em]
L2.ita\_t    &       0.490         \\
            &      (1.26)         \\
[1em]
\_cons      &      0.0290         \\
            &      (1.55)         \\
\hline
ita\_t       &                     \\
L.ita\_r     &       1.984\sym{*}  \\
            &      (2.13)         \\
[1em]
L2.ita\_r    &      -1.557         \\
            &     (-1.31)         \\
[1em]
L.ita\_t     &      -0.155         \\
            &     (-0.29)         \\
[1em]
L2.ita\_t    &     -0.0482         \\
            &     (-0.07)         \\
[1em]
\_cons      &      0.0770\sym{*}  \\
            &      (2.20)         \\
\hline
\(N\)       &           7         \\
\hline\hline
\multicolumn{2}{l}{\footnotesize \textit{t} statistics in parentheses}\\
\multicolumn{2}{l}{\footnotesize \sym{*} \(p<0.05\), \sym{**} \(p<0.01\), \sym{***} \(p<0.001\)}\\
\end{tabular}
}
\end{table}
===============================================

\begin{table}
	\label{var:lva}
\caption{VAR estimation: Latvia (LVA)}
\centering
{
\def\sym#1{\ifmmode^{#1}\else\(^{#1}\)\fi}
\begin{tabular}{l*{1}{c}}
\hline\hline
            &\multicolumn{1}{c}{(1)}\\
            &\multicolumn{1}{c}{lva\_r}\\
\hline
lva\_r       &                     \\
L.lva\_r     &      -0.419\sym{*}  \\
            &     (-1.97)         \\
[1em]
L2.lva\_r    &      -1.266\sym{***}\\
            &     (-3.53)         \\
[1em]
L.lva\_t     &       34.25\sym{***}\\
            &     (10.35)         \\
[1em]
L2.lva\_t    &       28.37\sym{***}\\
            &      (3.31)         \\
[1em]
\_cons      &     -0.0555\sym{***}\\
            &     (-7.86)         \\
\hline
lva\_t       &                     \\
L.lva\_r     &      0.0247         \\
            &      (1.41)         \\
[1em]
L2.lva\_r    &     -0.0392         \\
            &     (-1.32)         \\
[1em]
L.lva\_t     &       0.903\sym{***}\\
            &      (3.30)         \\
[1em]
L2.lva\_t    &      -0.932         \\
            &     (-1.32)         \\
[1em]
\_cons      &     0.00130\sym{*}  \\
            &      (2.23)         \\
\hline
\(N\)       &           7         \\
\hline\hline
\multicolumn{2}{l}{\footnotesize \textit{t} statistics in parentheses}\\
\multicolumn{2}{l}{\footnotesize \sym{*} \(p<0.05\), \sym{**} \(p<0.01\), \sym{***} \(p<0.001\)}\\
\end{tabular}
}
\end{table}
===============================================
\begin{table}
	\label{var:hun}
\caption{VAR estimation: Hungary (HUN)}
\centering
{
\def\sym#1{\ifmmode^{#1}\else\(^{#1}\)\fi}
\begin{tabular}{l*{1}{c}}
\hline\hline
            &\multicolumn{1}{c}{(1)}\\
            &\multicolumn{1}{c}{hun\_r}\\
\hline
hun\_r       &                     \\
L.hun\_r     &      -0.168         \\
            &     (-0.91)         \\
[1em]
L2.hun\_r    &      -0.251         \\
            &     (-1.59)         \\
[1em]
L.hun\_t     &       5.866\sym{**} \\
            &      (3.11)         \\
[1em]
L2.hun\_t    &      -0.809         \\
            &     (-0.46)         \\
[1em]
\_cons      &     -0.0360\sym{*}  \\
            &     (-1.98)         \\
\hline
hun\_t       &                     \\
L.hun\_r     &      0.0363         \\
            &      (0.86)         \\
[1em]
L2.hun\_r    &     -0.0715         \\
            &     (-1.96)         \\
[1em]
L.hun\_t     &       1.235\sym{**} \\
            &      (2.84)         \\
[1em]
L2.hun\_t    &      -0.852\sym{*}  \\
            &     (-2.10)         \\
[1em]
\_cons      &      0.0134\sym{**} \\
            &      (3.21)         \\
\hline
\(N\)       &           7         \\
\hline\hline
\multicolumn{2}{l}{\footnotesize \textit{t} statistics in parentheses}\\
\multicolumn{2}{l}{\footnotesize \sym{*} \(p<0.05\), \sym{**} \(p<0.01\), \sym{***} \(p<0.001\)}\\
\end{tabular}
}
\end{table}
===============================================

\begin{table}
	\label{var:nld}
\caption{VAR estimation: Netherlands (NLD)}
\centering
{
\def\sym#1{\ifmmode^{#1}\else\(^{#1}\)\fi}
\begin{tabular}{l*{1}{c}}
\hline\hline
            &\multicolumn{1}{c}{(1)}\\
            &\multicolumn{1}{c}{nld\_r}\\
\hline
nld\_r       &                     \\
L.nld\_r     &       0.812         \\
            &      (1.40)         \\
[1em]
L2.nld\_r    &      -0.389         \\
            &     (-0.91)         \\
[1em]
L.nld\_t     &      -1.601\sym{**} \\
            &     (-2.69)         \\
[1em]
L2.nld\_t    &       1.639         \\
            &      (1.89)         \\
[1em]
\_cons      &      0.0422         \\
            &      (0.19)         \\
\hline
nld\_t       &                     \\
L.nld\_r     &      -0.679\sym{*}  \\
            &     (-2.19)         \\
[1em]
L2.nld\_r    &       0.517\sym{*}  \\
            &      (2.26)         \\
[1em]
L.nld\_t     &       0.816\sym{*}  \\
            &      (2.56)         \\
[1em]
L2.nld\_t    &      -0.441         \\
            &     (-0.95)         \\
[1em]
\_cons      &      0.0826         \\
            &      (0.71)         \\
\hline
\(N\)       &           7         \\
\hline\hline
\multicolumn{2}{l}{\footnotesize \textit{t} statistics in parentheses}\\
\multicolumn{2}{l}{\footnotesize \sym{*} \(p<0.05\), \sym{**} \(p<0.01\), \sym{***} \(p<0.001\)}\\
\end{tabular}
}
\end{table}

\begin{table}
	\label{var:aut}
\caption{VAR estimation: Austria (AUT)}
\centering
{
\def\sym#1{\ifmmode^{#1}\else\(^{#1}\)\fi}
\begin{tabular}{l*{1}{c}}
\hline\hline
            &\multicolumn{1}{c}{(1)}\\
            &\multicolumn{1}{c}{aut\_r}\\
\hline
aut\_r       &                     \\
L.aut\_r     &      -0.175         \\
            &     (-0.46)         \\
[1em]
L2.aut\_r    &     -0.0233         \\
            &     (-0.08)         \\
[1em]
L.aut\_t     &       3.394         \\
            &      (1.52)         \\
[1em]
L2.aut\_t    &       3.274         \\
            &      (1.38)         \\
[1em]
\_cons      &      -0.216         \\
            &     (-1.87)         \\
\hline
aut\_t       &                     \\
L.aut\_r     &    -0.00510         \\
            &     (-0.09)         \\
[1em]
L2.aut\_r    &      0.0294         \\
            &      (0.65)         \\
[1em]
L.aut\_t     &      0.0775         \\
            &      (0.22)         \\
[1em]
L2.aut\_t    &       0.119         \\
            &      (0.32)         \\
[1em]
\_cons      &      0.0340         \\
            &      (1.90)         \\
\hline
\(N\)       &           7         \\
\hline\hline
\multicolumn{2}{l}{\footnotesize \textit{t} statistics in parentheses}\\
\multicolumn{2}{l}{\footnotesize \sym{*} \(p<0.05\), \sym{**} \(p<0.01\), \sym{***} \(p<0.001\)}\\
\end{tabular}
}
\end{table}
===============================================

\begin{table}
	\label{var:pol}
\caption{VAR estimation: Poland (POL)}
\centering
{
\def\sym#1{\ifmmode^{#1}\else\(^{#1}\)\fi}
\begin{tabular}{l*{1}{c}}
\hline\hline
            &\multicolumn{1}{c}{(1)}\\
            &\multicolumn{1}{c}{pol\_r}\\
\hline
pol\_r       &                     \\
L.pol\_r     &      -0.279         \\
            &     (-0.79)         \\
[1em]
L2.pol\_r    &      -0.167         \\
            &     (-0.69)         \\
[1em]
L.pol\_t     &       2.236\sym{**} \\
            &      (3.26)         \\
[1em]
L2.pol\_t    &      -1.314         \\
            &     (-1.77)         \\
[1em]
\_cons      &      0.0368\sym{***}\\
            &      (3.69)         \\
\hline
pol\_t       &                     \\
L.pol\_r     &      0.0667         \\
            &      (0.32)         \\
[1em]
L2.pol\_r    &      -0.113         \\
            &     (-0.80)         \\
[1em]
L.pol\_t     &       1.704\sym{***}\\
            &      (4.26)         \\
[1em]
L2.pol\_t    &      -0.836         \\
            &     (-1.93)         \\
[1em]
\_cons      &     0.00649         \\
            &      (1.12)         \\
\hline
\(N\)       &           7         \\
\hline\hline
\multicolumn{2}{l}{\footnotesize \textit{t} statistics in parentheses}\\
\multicolumn{2}{l}{\footnotesize \sym{*} \(p<0.05\), \sym{**} \(p<0.01\), \sym{***} \(p<0.001\)}\\
\end{tabular}
}
\end{table}
===============================================

\begin{table}
	\label{var:prt}
\caption{VAR estimation: Portugal (PRT)}
\centering
{
\def\sym#1{\ifmmode^{#1}\else\(^{#1}\)\fi}
\begin{tabular}{l*{1}{c}}
\hline\hline
            &\multicolumn{1}{c}{(1)}\\
            &\multicolumn{1}{c}{prt\_r}\\
\hline
prt\_r       &                     \\
L.prt\_r     &       0.583         \\
            &      (1.82)         \\
[1em]
L2.prt\_r    &      -0.157         \\
            &     (-0.42)         \\
[1em]
L.prt\_t     &       0.569         \\
            &      (0.68)         \\
[1em]
L2.prt\_t    &      -1.598         \\
            &     (-1.56)         \\
[1em]
\_cons      &      0.0481\sym{*}  \\
            &      (2.44)         \\
\hline
prt\_t       &                     \\
L.prt\_r     &     -0.0133         \\
            &     (-0.18)         \\
[1em]
L2.prt\_r    &       0.352\sym{***}\\
            &      (4.05)         \\
[1em]
L.prt\_t     &      0.0357         \\
            &      (0.18)         \\
[1em]
L2.prt\_t    &      -0.537\sym{*}  \\
            &     (-2.23)         \\
[1em]
\_cons      &     0.00426         \\
            &      (0.92)         \\
\hline
\(N\)       &           7         \\
\hline\hline
\multicolumn{2}{l}{\footnotesize \textit{t} statistics in parentheses}\\
\multicolumn{2}{l}{\footnotesize \sym{*} \(p<0.05\), \sym{**} \(p<0.01\), \sym{***} \(p<0.001\)}\\
\end{tabular}
}
\end{table}

\begin{table}
	\label{var:rou}
\caption{VAR estimation: Romania (ROU)}
\centering
{
\def\sym#1{\ifmmode^{#1}\else\(^{#1}\)\fi}
\begin{tabular}{l*{1}{c}}
\hline\hline
            &\multicolumn{1}{c}{(1)}\\
            &\multicolumn{1}{c}{rou\_r}\\
\hline
rou\_r       &                     \\
L.rou\_r     &       0.622\sym{*}  \\
            &      (2.27)         \\
[1em]
L2.rou\_r    &      -0.932\sym{*}  \\
            &     (-2.27)         \\
[1em]
L.rou\_t     &       96.17\sym{***}\\
            &      (3.76)         \\
[1em]
L2.rou\_t    &      -92.30\sym{***}\\
            &     (-3.47)         \\
[1em]
\_cons      &     -0.0781\sym{***}\\
            &     (-4.47)         \\
\hline
rou\_t       &                     \\
L.rou\_r     &   -0.000420         \\
            &     (-0.17)         \\
[1em]
L2.rou\_r    &     -0.0526\sym{***}\\
            &    (-14.41)         \\
[1em]
L.rou\_t     &       2.088\sym{***}\\
            &      (9.18)         \\
[1em]
L2.rou\_t    &      -1.027\sym{***}\\
            &     (-4.34)         \\
[1em]
\_cons      &   -0.000132         \\
            &     (-0.85)         \\
\hline
\(N\)       &           7         \\
\hline\hline
\multicolumn{2}{l}{\footnotesize \textit{t} statistics in parentheses}\\
\multicolumn{2}{l}{\footnotesize \sym{*} \(p<0.05\), \sym{**} \(p<0.01\), \sym{***} \(p<0.001\)}\\
\end{tabular}
}
\end{table}

\begin{table}
	\label{var:svk}
\caption{VAR estimation: Slovakia (SVK)}
\centering
{
\def\sym#1{\ifmmode^{#1}\else\(^{#1}\)\fi}
\begin{tabular}{l*{1}{c}}
\hline\hline
            &\multicolumn{1}{c}{(1)}\\
            &\multicolumn{1}{c}{svk\_r}\\
\hline
svk\_r       &                     \\
L.svk\_r     &      -0.912\sym{*}  \\
            &     (-2.33)         \\
[1em]
L2.svk\_r    &      -0.785\sym{*}  \\
            &     (-2.32)         \\
[1em]
L.svk\_t     &      -1.018         \\
            &     (-1.02)         \\
[1em]
L2.svk\_t    &       1.074         \\
            &      (0.97)         \\
[1em]
\_cons      &      0.0150\sym{**} \\
            &      (3.09)         \\
\hline
svk\_t       &                     \\
L.svk\_r     &       0.175         \\
            &      (1.16)         \\
[1em]
L2.svk\_r    &      0.0350         \\
            &      (0.27)         \\
[1em]
L.svk\_t     &       0.361         \\
            &      (0.94)         \\
[1em]
L2.svk\_t    &       0.553         \\
            &      (1.30)         \\
[1em]
\_cons      &     0.00123         \\
            &      (0.66)         \\
\hline
\(N\)       &           7         \\
\hline\hline
\multicolumn{2}{l}{\footnotesize \textit{t} statistics in parentheses}\\
\multicolumn{2}{l}{\footnotesize \sym{*} \(p<0.05\), \sym{**} \(p<0.01\), \sym{***} \(p<0.001\)}\\
\end{tabular}
}
\end{table}
===============================================

\begin{table}
\label{var:swe}
\caption{VAR estimation: Sweden (SWE)}
\centering
{
\def\sym#1{\ifmmode^{#1}\else\(^{#1}\)\fi}
\begin{tabular}{l*{1}{c}}
\hline\hline
            &\multicolumn{1}{c}{(1)}\\
            &\multicolumn{1}{c}{swe\_r}\\
\hline
swe\_r       &                     \\
L.swe\_r     &      -0.378         \\
            &     (-1.49)         \\
[1em]
L2.swe\_r    &     -0.0566         \\
            &     (-0.31)         \\
[1em]
L.swe\_t     &      -4.045\sym{***}\\
            &     (-4.97)         \\
[1em]
L2.swe\_t    &      -6.569\sym{***}\\
            &     (-6.86)         \\
[1em]
\_cons      &       0.388\sym{***}\\
            &      (7.22)         \\
\hline
swe\_t       &                     \\
L.swe\_r     &      0.0679         \\
            &      (0.32)         \\
[1em]
L2.swe\_r    &       0.132         \\
            &      (0.87)         \\
[1em]
L.swe\_t     &       0.469         \\
            &      (0.68)         \\
[1em]
L2.swe\_t    &      0.0514         \\
            &      (0.06)         \\
[1em]
\_cons      &    -0.00179         \\
            &     (-0.04)         \\
\hline
\(N\)       &           7         \\
\hline\hline
\multicolumn{2}{l}{\footnotesize \textit{t} statistics in parentheses}\\
\multicolumn{2}{l}{\footnotesize \sym{*} \(p<0.05\), \sym{**} \(p<0.01\), \sym{***} \(p<0.001\)}\\
\end{tabular}
}
\end{table}

\begin{table}
\caption{VAR estimation: UK (GBR)}
\centering
{
\def\sym#1{\ifmmode^{#1}\else\(^{#1}\)\fi}
\begin{tabular}{l*{1}{c}}
\hline\hline
            &\multicolumn{1}{c}{(1)}\\
            &\multicolumn{1}{c}{gbr\_r}\\
\hline
gbr\_r       &                     \\
L.gbr\_r     &     -0.0660         \\
            &     (-0.17)         \\
[1em]
L2.gbr\_r    &      0.0609         \\
            &      (0.15)         \\
[1em]
L.gbr\_t     &       0.505         \\
            &      (1.26)         \\
[1em]
L2.gbr\_t    &       0.167         \\
            &      (0.37)         \\
[1em]
\_cons      &     0.00631         \\
            &      (0.22)         \\
\hline
gbr\_t       &                     \\
L.gbr\_r     &       0.878\sym{***}\\
            &      (3.70)         \\
[1em]
L2.gbr\_r    &      -0.152         \\
            &     (-0.60)         \\
[1em]
L.gbr\_t     &       0.548\sym{*}  \\
            &      (2.25)         \\
[1em]
L2.gbr\_t    &      -0.126         \\
            &     (-0.46)         \\
[1em]
\_cons      &  -0.0000736         \\
            &     (-0.00)         \\
\hline
\(N\)       &           7         \\
\hline\hline
\multicolumn{2}{l}{\footnotesize \textit{t} statistics in parentheses}\\
\multicolumn{2}{l}{\footnotesize \sym{*} \(p<0.05\), \sym{**} \(p<0.01\), \sym{***} \(p<0.001\)}\\
\end{tabular}
}
\label{var:uk}
\end{table}


\begin{thebibliography}{10}
\providecommand{\url}[1]{{#1}}
\providecommand{\urlprefix}{URL }
\expandafter\ifx\csname urlstyle\endcsname\relax
  \providecommand{\doi}[1]{DOI~\discretionary{}{}{}#1}\else
  \providecommand{\doi}{DOI~\discretionary{}{}{}\begingroup
  \urlstyle{rm}\Url}\fi

\bibitem{ECI}
Economic complexity index database. atlas of economic complexity. (2017).
\newblock \urlprefix\url{http://atlas.cid.harvard.edu/rankings/country/}

\bibitem{GDP}
Gdp (per capita by country) database. knoema world data atlas . (2017).
\newblock
  \urlprefix\url{https://knoema.com/pjeqzh/gdp-per-capita-by-country-1980-2014}

\bibitem{Kenett_16}
Bookstaber, R., Kenett, D.Y.: Looking deeper, seeing more: a multilayer map of
  the financial system.
\newblock Available at https://financialresearch.gov/briefs/2016/
  07/14/multilayer-map  (2016)

\bibitem{Fagiolo_JEIC_13a}
Duenas, M., Fagiolo, G.: Modeling the international-trade network: A gravity
  approach.
\newblock Journal of Economic Interaction and Coordination \textbf{8}, 155--178
  (2013)

\bibitem{Hidalgo_2009}
Hidalgo, C.A., Hausmann, R.: The building blocks of economic complexity.
\newblock Proceedings of the national academy of sciences \textbf{106}(26),
  10,570--10,575 (2009)

\bibitem{Lee_2002}
Lee, C.S.: The linkage between the fdi and trade of china, japan and korea: the
  korean perspective.
\newblock In: Seoul: Korean Institute for International Economic Policy. Paper
  prepared for presentation at the DRC/NIRA/KIEP symposium on Strengthening
  Economic Cooperation in Northeast Asia: Facilitating Investment between
  China, Japan and Korea held in Beijing on September, vol.~29, p. 2002 (2002)

\bibitem{Lee_2016}
Lee, K.M., Goh, K.I.: Strength of weak layers in cascading failures on
  multiplex networks: case of the international trade network.
\newblock Scientific reports \textbf{6}, 26,346 (2016)

\bibitem{Lee_2015}
Lee, K.M., Min, B., Goh, K.I.: Towards real-world complexity: an introduction
  to multiplex networks.
\newblock The European Physical Journal B \textbf{88}(2), 48 (2015)

\bibitem{Stanley_1998}
Lee, Y., Amaral, L.A.N., Canning, D., Meyer, M., Stanley, H.E.: Universal
  features in the growth dynamics of complex organizations.
\newblock Physical Review Letters \textbf{81}(15), 3275 (1998)

\bibitem{Liu_1998}
Liu, L., Graham, E.: The Relationship Between Trade and Foreign Investment:
  Empirical Results for Taiwan and South Korea.
\newblock Working paper series (Institute for International Economics (U.S.))).
  Institute for International Economics (1998).
\newblock \urlprefix\url{https://books.google.co.in/books?id=rIAUnQAACAAJ}

\bibitem{Qadan_15}
Qadan, M., Yagil, J.: International co-movements of real and financial economic
  variables.
\newblock Applied Economics \textbf{47}(31), 3347--3366 (2015)

\bibitem{Sharma_2017_b}
Sharma, K., Gopalakrishnan, B., Chakrabarti, A.S., Chakraborti, A.: Financial
  fluctuations anchored to economic fundamentals: A mesoscopic network
  approach.
\newblock Scientific reports \textbf{7}(1), 8055 (2017)

\bibitem{Fagiolo_JEIC_13b}
Squartini, T., Fagiolo, G., Garlaschelli, D.: Null models of economic networks:
  The case of the world trade web.
\newblock Journal of Economic Interaction and Coordination \textbf{8}, 75--107
  (2013)

\bibitem{Thompson_reuters}
{Thompson Reuters Eikon Database}:  (2016).
\newblock
  \urlprefix\url{https://customers.thomsonreuters.com/eikon/index.html}.
\newblock Accessed on 9th November, 2016

\bibitem{Wang_2018}
Wang, G.J., Xie, C., Stanley, H.E.: Correlation structure and evolution of
  world stock markets: Evidence from pearson and partial correlation-based
  networks.
\newblock Computational Economics \textbf{51}(3), 607--635 (2018)


\bibitem{Vodenska_2015}
Curme, C., Stanley, H. E. and Vodenska, I.: Coupled network approach to predictability of financial market returns and news sentiments.
\newblock International Journal of Theoretical and Applied Finance  \textbf{18}(07), 1550043 (2015)

\end{thebibliography}
\end{document}